\documentclass[a4paper, 12pt]{article}
\usepackage{a4}
\usepackage{amsfonts}
\usepackage{amssymb}
\usepackage{epsfig}
\usepackage{psfig}

\newcommand{\be}{\begin{equation}}
\newcommand{\ee}{\end{equation}}
\newcommand{\bea}{\begin{eqnarray}}
\newcommand{\eea}{\end{eqnarray}}
\newcommand{\mbb}{\mathbb}
\newcommand{\ti}{\times}
\newcommand{\half}{\frac{1}{2}}

\begin{document}

\title{
\begin{flushright} \vspace{-2cm} 
{\small DAMTP-2004-42 \\ \vspace{-0.35cm} 
hep-th/0404254} \end{flushright}
\vspace{3cm}
Type IIA Orientifolds on General \\
Supersymmetric  $\mbb{Z}_N$ Orbifolds
}
\author{} 
\date{}

\maketitle

\begin{center}
Ralph Blumenhagen,\footnote{e-mail: R.Blumenhagen@damtp.cam.ac.uk} 
Joseph P. Conlon \footnote{e-mail: J.P.Conlon@damtp.cam.ac.uk} and  
Kerim Suruliz \footnote{e-mail: K.Suruliz@damtp.cam.ac.uk} \\
\vspace{0.5cm}
\emph{DAMTP, Centre for Mathematical Sciences,} \\
\emph{Wilberforce Road, Cambridge, CB3 0WA, UK} \\
\vspace{2cm}
\end{center}

\begin{abstract}
\noindent
We construct Type IIA orientifolds for general supersymmetric 
$\mbb{Z}_N$ orbifolds. In particular, we provide the methods
to deal with the non-factoris\-able six-dimensional tori for the
cases $\mbb{Z}_7$, $\mbb{Z}_8$, $\mbb{Z}'_8$, $\mbb{Z}_{12}$
and $\mbb{Z}'_{12}$. As an application of these methods
we explicitly construct many new orientifold models.

\end{abstract}

\thispagestyle{empty}
\clearpage

\tableofcontents

\section{Introduction}

The last years have seen some considerable effort in constructing new
string vacua with D-branes in the background. The natural set-up
involves so-called orientifold models, for which tadpole cancellation
really forces us to introduce D-branes, which
from the low energy point of view 
extends the closed string gravity theory by gauge degrees of freedom
(see \cite{as02} and refs. therein).
Therefore, besides the heterotic string these orientifold models
constitute a class of string backgrounds which exhibit 
interesting phenomenological properties. 

It is by now well established that chirality can be implemented
in Type IIA orientifold models by using intersecting D6-branes,
where each topological intersection point gives rise to a chiral fermion.
Indeed, following some earlier work 
\cite{CB95,bdl96,bgk99,ab99,bgk99a,GP99,fhs00,bgkl00,bgkl00a,aads00} 
many of these so-called
intersecting D-brane models have been constructed so far, 
which feature some of the properties of the non-supersymmetric
\cite{afiru00,afira00a,bkl00,imr01,bklo01,bkl01,Ck02}
or supersymmetric Standard Model \cite{csu01,csu01a,bgo02,GH03,cp03,LG03,ho04,cll04}
(please consult  the reviews \cite{AU03,DL04} for more refs.). 

However, the class of closed string backgrounds studied so far
is still quite limited. There are two principal classes of models
which are still under investigation. 
One is the class of toroidal orbifold backgrounds, for which only a limited
number of examples have been investigated so far 
\cite{csu01,csu01a,bgo02,GH03,LG03,ho04}. The second class
are orientifolds of Gepner models \cite{abp96,bw98}, 
for which the general structure
of the one-loop amplitudes and the tadpole cancellation
conditions were worked out  recently 
in \cite{aaln03,RB03a,bhhw04,bw04,dhs04,bw04a,aaj04}.

The aim of this paper is  to extend the work on Type IIA
orientifolds of toroidal  orbifolds,
where so far only the cases $\mbb{Z}_2\times \mbb{Z}_2$, $\mbb{Z}_3$,
$\mbb{Z}_4$, $\mbb{Z}_4\times \mbb{Z}_2$ and $\mbb{Z}_6$ have been studied
to the extent that chiral intersecting D6-brane models have been
constructed. 
The main reason for focusing  on these orbifolds is
of technical nature, namely that in these cases
the complex structure of the six-dimensional torus
can be chosen such that it decomposes as
$T^6=T^2\times T^2\times T^2$. All the other $\mbb{Z}_N$
Type IIA orientifolds, namely $\mbb{Z}_7$, $\mbb{Z}_8$, $\mbb{Z}'_8$
$\mbb{Z}_{12}$ and $\mbb{Z}'_{12}$, are not completely
factorisable in the above sense and so far remained largely
unexplored. It is the aim of this paper to resolve
some of the technical problems with describing
these orientifolds properly and to provide the necessary 
technical tools. 

Note, that Type IIB orientifolds on these
non-factorisable orbifolds are technically much simpler
and partition functions could be computed fairly 
straightforwardly \cite{ks97,afiv98}.
However, it was found there that certain Type IIB $\mbb{Z}_N$
orientifolds with $N$ even
do not allow for tadpole cancelling configurations of 
$D9$ and $D5$ branes \cite{zwart,afiv98,kst98}. 
We would like to point out that first the Type IIA models
considered here are not T-dual to the Type IIB models
mentioned above and second that in our case we always find
tadpole cancelling configurations. This  is surely related to the
fact that here only untwisted and almost trivial $\mbb{Z}_2$
twisted sector tadpoles appear, whereas in the Type IIB case
twisted tadpoles in all sectors do arise. 

Of course we are finally interested in constructing chiral
supersymmetric  intersecting brane models on these
non-factorisable orbifolds. After determining
the general form of the Klein-bottle amplitudes, as a starting point
we restrict ourselves to the solutions
with D6-branes placed on top of the orientifold
planes, thus generalising the work of \cite{bgk99a,GP99,fhs00}.  
As we will discuss, to construct these solutions properly,
some new ingredients in the computation of the one-loop
amplitudes need to be employed. 

This paper is organised as follows. In section 2 we first review
the classification of supersymmetric $\mbb{Z}_N$ orbifolds
allowing for a crystallographic action of the symmetry. 
Then we provide  the general framework to deal with the
non-factorisable orbifolds and derive  general results
for the various one-loop amplitudes. Section 3 contains
the rules for computing  both the closed and the
open string spectrum. In section 4 we revisit and  extend
the factorisable orbifolds  studied already in \cite{bgk99a}. 
In section 5 we apply our techniques to the construction
of new orientifolds on non-factorisable orientifolds, where
we discuss quite a large number in detail and encounter some
new technical subtleties in computing the one-loop amplitudes.
As a first step, here we focus on the non-chiral solutions to
the tadpole cancellation conditions with D6-branes right
on top of the orientifold planes. 
Finally, section 6 contains our conclusions.

\section{Type IIA orientifolds on non-factorisable orbifolds}
\label{Amplitudes}

In this section we first review some facts about $\Omega R$ orientifolds
and the classification of supersymmetric six-dimensional orbifolds. 
In the second part we develop new methods to deal with the
non-factorisable lattices and to compute one-loop
partition functions from which one can extract the
tadpole cancellation conditions.

\subsection{Definition of $\Omega R$ orientifolds}
\label{definitions}
Orientifolds are a natural method for introducing D-branes into a
theory. Suppose we start with type II string theory on a background
$\mbb{M}^4 \times X_6$, where for our purposes $X_6$ will be assumed compact.
An orientifold is defined by taking the quotient of this theory
by $H=F + \Omega G$, where $F$ and $G$ are discrete groups of target
space symmetries. If $G$ were empty, we would have an orbifold - the
orientifold is defined by the fact that the symmetry we divide out by involves worldsheet
orientation reversal $\Omega$. We then project onto states that are
invariant under this symmetry.

The fixed points of $G$ define an O-plane. These are non-dynamical,
geometric surfaces that carry R-R charge and have a nonzero tadpole amplitude to
emit closed strings into the bulk. Unoriented closed string theories
are generally inconsistent due to this uncancelled R-R
charge located on the O-planes. In the compact space $X_6$, the R-R flux
has nowhere to escape to and it is necessary
to add D-branes to cancel the overall R-R charge. 
Closed strings then couple to both D-branes and O-planes, each giving
a tadpole amplitude for the emission of closed strings into the
vacuum. 
The disc (D-brane) and $\mathbb{RP}_2$ (O-plane) tadpoles give 
rise to infrared divergences due to closed string exchange in the one-loop
diagrams. 
This can be computed as a 
tree channel diagram, by constructing boundary and crosscap states and finding their overlap
(for a review see \cite{Gaberdiel}). 
However, world-sheet duality means that we can reinterpret
closed string tree channel diagrams
as open string or closed string one loop diagrams. 
In this paper we are not making explicit use of the
tree channel approach, but instead start directly with the computation of the 
relevant loop channel amplitudes, i.e.
the Klein bottle, annulus and M\"obius strip amplitude.

In this paper we will consider what are called $\Omega R$
orientifolds. The group $G$ always includes the element $R$, which
acts as
\bea
R: Z_i \leftrightarrow \bar{Z}_i & & i = 1,2,3
\eea
on the complex coordinates
$Z_1 = x_4 + ix_5$, $Z_2 = x_6 + ix_7$, $Z_3 = x_8 + ix_9$. 
The action of $\Omega R$ on R-R states is then
\be
\Omega R |s_0, s_1, s_2, s_3 \rangle \otimes |s'_0, s'_1, s'_2,
s'_3 \rangle = -|s'_0, -s'_1, -s'_2, -s'_3 \rangle \otimes
|s_0, -s_1, -s_2, -s_3 \rangle
\ee
and consequently $\Omega R$ is, for four-dimensional
compactifications, a symmetry of type IIA string theory. Note that for
the analogous six-dimensional case $\Omega R$ is a symmetry of type IIB
\footnote{Note, the action of $\Omega R$ on the R-R states guarantees that 
the orientifold is T-dual to the Type IIB $\Omega$ orientifold and therefore
preserves supersymmetry. Therefore, the action implicitly includes
the $(-1)^{F_L}$ factor found by A.Sen \cite{sen,sena}. }.

The models we consider are $\Omega R$ orientifolds of type IIA orbifolds.
It was shown in the mid eighties  \cite{dhvwa,dhvwb} that all orbifold actions $\Theta$
satisfying
modular invariance and preserving
4-dimensional $N=2$ supersymmetry
can be written as
\begin{eqnarray}
Z_i &\to& \exp(2\pi i v_i) Z_i \nonumber \\
\bar{Z}_i &\to& \exp(-2 \pi i v_i) \bar{Z}_i
\end{eqnarray}
where $v_i$ has the possible values given in table \ref{orbifoldstable}.

\begin{table}
\caption{Possible orbifold actions preserving N=1 supersymmetry}
\centering
\vspace{3mm}
\label{orbifoldstable}
\begin{tabular}{|c|c|c|c|c|c|}
\hline
$\mathbb{Z}_3$ & $(1,1,-2)/3$ & $\mathbb{Z}_4$ & $(1,1,-2)/4$ & $\mathbb{Z}_6$
&
$(1,1,-2)/6$ \\
$\mathbb{Z}_6^{'}$ & $(1,2,-3)/6$ & $\mathbb{Z}_7$ & $(1,2,-3)/7$ &
$\mathbb{Z}_8$ &
$(1,3,-4)/8$ \\
$\mathbb{Z}_8^{'}$ & $(2,1,-3)/8$ & $\mathbb{Z}_{12}$ & $(4,1,-5)/12$ &
$\mathbb{Z}_{12}^{'}$ & $(-6,1,5)/12$ \\
\hline
\end{tabular}
\end{table}

The requirement that $\Theta$ act crystallographically places stringent
conditions on the compact space $X_6$, namely $X_6$ must be a toroidal
lattice and in fact there are only 18 distinct possibilities 
\cite{Erler}. These are listed in table \ref{lattices} together
with the corresponding numbers of $(1,1)$ and $(1,2)$-forms, $h^{1,1}$ and
$h^{1,2}.$
For 15 of these cases, the orbifold action can be realised as the Coxeter element
$\omega = \Gamma_1 \Gamma_2 \Gamma_3 \Gamma_4 \Gamma_5 \Gamma_6$
acting on the root lattice of an appropriate
Lie algebra. For the other three, the orbifold action is instead realised
as a combination of Weyl reflections and outer automorphisms acting on
the Lie algebra root lattice.
This is shown in the last column of table
\ref{lattices}. $\Gamma_i$ is a Weyl reflection on the simple root $i$ and
$P_{ij}$ exchanges roots $i$ and $j$. For most of the orbifolds in
this paper, and for all those we study in detail, there is no
distinction between the orbifold action $\Theta$ and the Coxeter
element $\omega$. However, notationally we will tend to use $\Theta$
when referring to an element of the orbifold group and $\omega$ when
referring to its action on the basis vectors of the lattice. We trust
this will not cause confusion. 

\begin{table}
\caption{The 18 symmetric $\mbb{Z}_N$ orbifolds}
\centering
\vspace{3mm}
\label{lattices}
\begin{tabular}{|cr|c|c|c|c|}
\hline
\multicolumn{2}{|c|}{ Case}
& Lie algebra root lattice & $h^{1,1}$ & $h^{1,2}$ & Orbifold action $\Theta$ \\
\hline
\hline
1 & $\mbb{Z}_3$ & $A_2 \times A_2 \times A_2$ & 36 & - & $\omega$\\
2 & $\mbb{Z}_4$ & $A_1 \ti A_1 \ti B_2 \ti B_2$ & 31 & 7 & $\omega$\\
3 & $\mbb{Z}_4$ & $A_1 \ti A_3 \ti B_2$ & 27 & 3 & $\omega$\\
4 & $\mbb{Z}_4$ & $A_3 \ti A_3$ & 25 & 1 & $\omega$\\
5 & $\mbb{Z}_6$ & $A_2 \ti G_2 \ti G_2$ & 29 & 5 & $\omega$\\
6 & $\mbb{Z}_6$ & $G_2 \ti A_2 \ti A_2$ & 25 & 1 & $\Gamma_1 \Gamma_2 \Gamma_3 \Gamma_4 P_{36}
P_{45}$\\
7 & $\mbb{Z}_6'$ & $A_1 \ti A_1 \ti A_2 \ti G_2$ & 35 & 11 & $\omega$\\
8 & $\mbb{Z}_6'$ & $A_2 \ti D_4$ & 29 & 5 & $\omega$\\
9 & $\mbb{Z}_6'$ & $A_1 \ti A_1 \ti A_2 \ti A_2$ & 31 & 7 & $\Gamma_1 \Gamma_2 \Gamma_3 \Gamma_4
P_{36} P_{45}$\\
10 & $\mbb{Z}_6'$ & $A_1 \ti A_5$ & 25 & 1 & $\omega$\\
11 & $\mbb{Z}_7$ & $A_6$ & 24 & - & $\omega$ \\
12 & $\mbb{Z}_8$ & $B_4 \times D_2$ & 31 & 7 & $\omega$ \\
13 & $\mbb{Z}_8$ & $A_1 \times D_5$ & 27 & 3 & $\omega$ \\
14 & $\mbb{Z}_8'$ & $B_2 \ti B_4$ & 27 & 3 & $\omega$ \\
15 & $\mbb{Z}_8'$ & $A_3 \ti A_3$ & 24 & - & $\Gamma_1 \Gamma_2 \Gamma_3 P_{16} P_{25} P_{34}$
\\
16 & $\mbb{Z}_{12}$ & $A_2 \ti F_4$ & 29 & 5 & $\omega$ \\
17 & $\mbb{Z}_{12}$ & $E_6$ & 25 & 1 & $\omega$ \\
18 & $\mbb{Z}_{12}'$ & $D_2 \ti F_4$ & 31 & 7 & $\omega$ \\
\hline
\end{tabular}

\end{table}

So far $\Omega R$ orientifolds have only been constructed for a few of 
the orbifolds in table \ref{lattices}.
To construct the orientifold, $\Omega R$ must be a well-defined
symmetry of the theory
and thus $R$ must act crystallographically on the lattice.
The cases 1,2,5 and 7 were first studied in \cite{bgk99a} and have lattices
factorisable as $T^2 \ti T^2 \ti T^2$. For these cases, the action of $R$ is found more or less
by inspection. However, the generic lattice is non-factorisable and it is not obvious how to visualise
it.

\begin{figure}[ht]
\begin{center}
\makebox[10cm]{
\epsfxsize=15cm
\epsfysize=5cm
\epsfbox{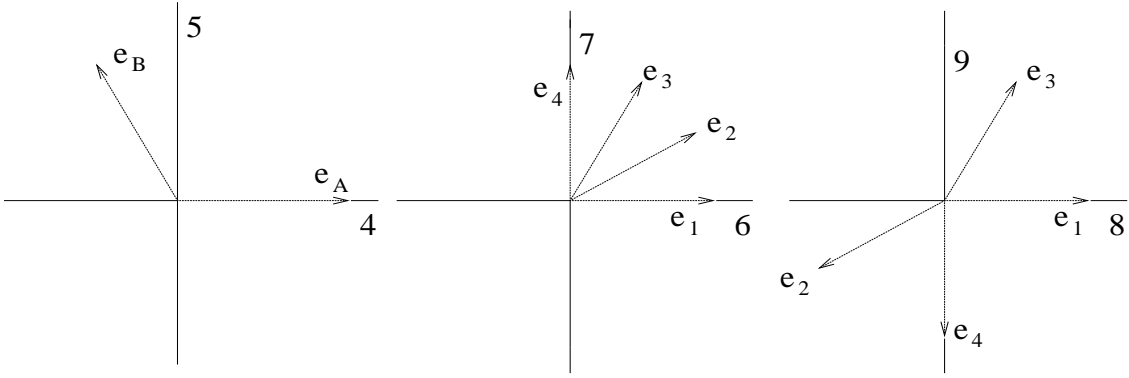}}
\end{center}
\caption{The $\mathbb{Z}_{12}$ lattice vectors. The arrow marked ${\bf e}_i$ in each plane is not ${\bf e}_i$ itself; rather, it is the
component of ${\bf e}_i$
in that plane. }
\label{z12latticepicture}
\end{figure}

To study the other cases in table \ref{lattices} we need to find a crystallographic implementation of $R$. 
The rotation planes of the orbifold action $\omega$ are
orthogonal. If we take a
given lattice vector, we can decompose it into components lying in each of
the
orbifold rotation planes. On each component, $\omega$ acts as a pure
rotation.
We choose one lattice vector, ${\bf e}_1$, of minimal size. For convenience, we
orient the rotation planes such that ${\bf e}_1$ lies along the $x$-axis in each
plane. As
$\omega$ acts crystallographically, $\omega {\bf e}_1$ is also a lattice vector.
Repeated action of $\omega$ generates a basis of lattice vectors. It is
necessary to check that this is
actually a basis for the lattice; in practice 
this is ensured by requiring ${\bf e}_1$ to be of
minimal size. Then, if $\omega^N = 1$, 
\be
R: \omega^{k}{\bf e}_i \leftrightarrow \omega^{N-k}{\bf e}_i
\ee
and is manifestly crystallographic. As an illustration of this construction, in figure \ref{z12latticepicture} 
the lattice vectors are shown for the $\mbb{Z}_{12}$ case, where the Lie algebra is $A_2 \times F_4$
and the orbifold rotation
$v = \frac{1}{12}(4,1,-5)$. 


%

Table \ref{lattices} lists the Lie algebras on whose root lattices the
orbifold action is implemented. However, we are not interested in the root
structure of the algebra
\emph{per se}, but rather in the lattice derived from it. Also, the simple roots
in general are of different
magnitude, but our construction generates basis vectors having the
same size. Thus the basis we use, although a perfectly good basis for the lattice, does not
 generally consist of root vectors. We
will see an example of this
in the $\mbb{Z}_8'$ case below. 

For the factorisable $T^2 \times T^2 \times T^2$ case, it was found that there are two distinct ways of
implementing the
$\Omega R$ projection in each $T^2$. These were called {\bf A} and
{\bf B} type lattices,
and
correspond to the
$R$ fixed plane being either along the root vectors or at a rotation of
$\omega^{\frac{1}{2}}$. One effect
of using more generic lattices is to reduce this freedom - the action of $R$ in
one rotation plane
determines its action in another.
For each independent torus - of whatever dimension - in
the lattice, there are two possible crystallographic actions of $\Omega R$.
As was observed  for the factorisable cases in \cite{bgk99} \cite{bgk99a}, in certain
models the tree channel amplitudes had the peculiar feature that the 
different contributions had the right prefactor to be interpreted
as a complete projector. In fact this feature served as a guiding principle
for model building and it was even claimed that it is necessary for consistency
of the model. This claim was corrected in \cite{bbkl02} where it was stated that
the other choices of the orientifold action also lead to consistent
models. In section 4 we will reconsider this issue for the factorisable
case and explicitly construct all the consistent models where the complete projector does not
appear.

We will implement the action of $\Omega R$ on the closed string modes as
\bea
\label{conventions}
(\Omega R) \psi_r (\Omega R)^{-1}& = & \tilde{\psi}_r \nonumber\\
(\Omega R) \tilde{\psi}_r (\Omega R)^{-1}& = & \psi_r
\eea
and likewise for the $\alpha_r$, $\tilde{\alpha}_r$ oscillators.
The loop channel amplitudes contributing to R-R exchange in tree channel are then
\bea
KB & = & 
\label{kb}
c \int_0^\infty \frac{dt}{t^3} \textrm{Tr}_{NSNS} 
\left( \Omega{\cal R} 
\frac{\left( 1+\Theta+ \cdots +\Theta^{N-1} \right)}{N} 
 e^{-2\pi t \left( L_0 + \bar{L}_0 \right) }
\right) \nonumber \\
A & = &
\label{annulus}
\frac{c}{4} \int_0^\infty \frac{dt}{t^3} \textrm{Tr}_{NS}
\left( (-1)^F \frac{\left( 1+\Theta+ \cdots +\Theta^{N-1} \right)}{N}
e^{-2\pi t L_0} \right) \nonumber \\
MS & = &
\label{ms}
-\frac{c}{4} \int_0^\infty \frac{dt}{t^3} \textrm{Tr}_{R}
\left( \Omega R \frac{\left( 1+\Theta+ \cdots +\Theta^{N-1} \right)}{N}
e^{-2\pi t L_0} \right)
\eea
Here $ c \equiv V_4 / (8\pi^2 \alpha')^2$ arises from the integration over non-compact momenta.
These amplitudes involve combinations of $\vartheta$ functions and lattice sums.
We transform these amplitudes to tree channel using
\be
t = \left\{ \begin{array}{ll} \frac{1}{2l} & \textrm{Annulus} \\ \frac{1}{4l} & \textrm{Klein Bottle} \\
\frac{1}{8l} & \textrm{M\"obius Strip} \end{array} \right.
\ee
As $R \Theta^k = \Theta^{n-k} R$, $(\Omega R \Theta^k)^2 = 1$ and so we expect only untwisted states to propagate
in tree channel.
The oscillator contributions in a given twisted sector are straightforward. We convert these to tree
channel using the modular transformation properties of the $\vartheta$ functions. These, together with the resulting tree channel
expressions, are given in the appendix. The lattice contributions are more involved and will be considered below.

In the Klein bottle trace, all insertions in the trace give rise to the same oscillator contribution. However,
insertions of the form $\Omega R \Theta^{2k}$ and $\Omega R \Theta^{2k+1}$ generically give rise to
different lattice contributions.
For the annulus and M\"obius sectors, the branes are placed at orientations related by $\Theta^\half$. `Twisted sector'
open string amplitudes then correspond to strings stretching between branes at angles. The only non-vanishing
insertions in the annulus amplitude are 1 and, if $N$ is even, $\Theta^{\frac{N}{2}}$. The latter insertion leaves
all the D6-branes invariant and hence induces an action on the Chan-Paton factors described by a matrix
$\gamma_{\frac{N}{2}}$.
The twisted tadpole cancellation condition then implies $\textrm{tr} (\gamma_{\frac{N}{2}}) = 0$. 
As this is the sole effect of this insertion, when we consider
particular models we will not write out explicitly the amplitudes arising from this term.

In the M\"obius amplitude, only ($6_i$, $6_{i+2 \lambda}$) strings 
can be invariant under an $\Omega R \Theta^k$ insertion and
thus contribute in the M\"obius strip amplitude. There are 
two insertions under which such strings are invariant, schematically
$\Omega R \Theta^\lambda$ and $\Omega R \Theta^{\lambda + \frac{N}{2}}$. These contribute to the
$\Theta^\lambda$ and $\Theta^{\lambda + \frac{N}{2}}$ twisted sectors. For both the annulus and M\"obius strip amplitudes, strings
starting on the $6_{2k}$ and $6_{1+2k}$ branes generically give different lattice contributions.

\subsection{Dealing with non-factorisable lattices}
\label{Lattice}

For a non-factorisable lattice it is not obvious how the lattice modes should be computed.
In sectors where there are fixed tori the lattice contributes momentum and
winding modes to the
partition function. 
If general momentum and winding modes can be written
\bea
{\bf p} &=& \sum n_i {\bf p}_i  \nonumber\\
{\bf w} &=& \sum m_i {\bf w}_i,\quad  m_i, n_i \in \mbb{Z}
\eea
then the loop channel partition function contains a sum
\be
\label{latticesum}
\sum_{n_i} \exp (-\delta \pi t n_i M_{ij} n_j)\sum_{m_i} \exp (-\delta \pi t m_i W_{ij} m_j)
\ee
where $n_i$, $m_i \in \mbb{Z}$, $M_{ij} = {\bf p}_i \cdot {\bf p}_j$, $W_{ij} = {\bf w}_i \cdot {\bf w}_j$ and 
$\delta = \left\{ \begin{array}{ll} 1 & \textrm{Klein bottle} \\ 2 & \textrm{Annulus, M\"obius strip} 
\end{array} \right. $ 
We then transform to tree channel using
\be
t = \left\{ \begin{array}{ll} \frac{1}{2l} & \textrm{Annulus} \\ \frac{1}{4l} & \textrm{Klein Bottle} \\
\frac{1}{8l} & \textrm{M\"obius Strip} \end{array} \right.
\ee
and the generalised Poisson resummation formula
\be
\sum_{n_i} \exp (-\pi t n_i A_{ij} n_j) = \frac{1}{t^{{\rm dim}(A)\over 2}
    (\det A)^{\frac{1}{2}}} \sum_{n_i}
\exp(-\frac{\pi}{t}n_i A^{-1}_{ij} n_j).
\ee

To find the momentum modes we need to find vectors in the dual lattice 
invariant under $\Omega R$.
The appropriate dual lattice differs in open and closed sectors. Closed strings can move freely throughout the
lattice and so the dual lattice is that of the $T^6$. Open strings are tied to a brane and the dual lattice is that
of the vectors spanning the brane. 
Winding modes are perpendicular to the
momentum modes. 
In the annulus case, all winding modes running from a brane
to itself contribute to the sum (\ref{latticesum}). In the Klein
bottle and M\"obius strip amplitudes, there is an insertion of $\Omega
R$ into the trace. In the M\"obius strip amplitudes, where there are
generically open string winding modes that are fractional multiples of lattice
vectors, the insertion of $\Omega R$ in the trace can cause the centre of mass
coordinate of the string to be shifted by
\be
T: {\bf x} \to {\bf x} + {\bf a} .
\ee
$T$ acts on momentum modes as
\be
T | {\bf p} \rangle = \exp(2 \pi i {\bf p} \cdot {\bf a}) | {\bf p}
\rangle.
\ee
In the particular case where ${\bf p} = \frac{n {\bf e}_x}{R}$ and
${\bf T}: {\bf x} \to {\bf x} + \frac{R}{2}{\bf e}_x$, the sum over
momentum modes in (\ref{latticesum}) is then
\be
\sum_{n_i} (-1)^n \exp (-\pi t n_i A_{ij} n_j).
\ee
Transformed to tree channel, this sum does not actually diverge in the
$l \to \infty$ limit and so such winding modes do not contribute to
the tadpoles.
Lattice modes also occur when one complex plane is left invariant under the
orbifold action, when an analogous version of the above discussion
applies.

The lattice also contributes to the amplitude through fixed points and brane
intersection numbers.
For a non-factorisable lattice, it is difficult to find these visually.
For the former, the Lefschetz fixed point theorem is not sufficient
as we also need to know which points are invariant under the action of $\Omega
R$. We therefore need to
compute explicitly the location of the fixed points using the action of
$\omega$. This is tedious but not
difficult. A useful reference for this and other details of the non-factorisable 
lattices is \cite{Casas}, although the lattice bases used there differ from ours.
To calculate brane intersection numbers we need to find a spanning
set of lattice vectors for each
of the branes. The intersection points are then given by solving a set of
equations of the
schematic form
\be
\sum \alpha_i {\bf v}_i = \sum \beta_i {\bf w}_i
\ee
for nontrivial (not in $\mathbb{Z}$) values of $\alpha_i$. In the case of the
M\"obius strip, we also need
to investigate whether the intersection points are invariant under $\Omega R
\Theta^k$ for the
appropriate value of $k$.

Actually, for the annulus amplitude we would like to present a simple
method for calculating the contribution of the lattice
modes and the intersection numbers. Let ${\bf e}_i$ be the basis vectors of the lattice and let 
$g_{ij} = {\bf e}_i \cdot {\bf e}_j$. Let ${\bf v}_i, i=1,2,3$ be the lattice vectors that describe the 3-cycle
wrapped by the brane, such that any point $\bf x$ on the brane can be uniquely written
\bea
\label{branewrap}
{\bf x} = \sum \alpha_i {\bf v}_i & & \alpha_i \in [0,1).
\eea
We then define $(M_A)_{ij} = {\bf v}_i \cdot {\bf v}_j$. The lattice dual to the brane is now
given by ${\bf v}_i^* = {\bf v}_j (M_A^{-1})_{ji}$, with
${\bf v}_i^* \cdot {\bf v}_j^* = (M_A^{-1})_{ij}$. The contribution of momentum modes to the
one-loop amplitude is then
\be
\label{momcont}
\sum_{n_i} \exp \left( -2\pi t n_i (M_A^{-1})_{ij} n_j \right) \to l^{{\dim(M_A)\over2}}(\det M_A)^{\half} \sum_{n_i} 
\exp \left( -\pi l n_i (M_A)_{ij} n_j \right).
\ee
Suppose a generic winding mode is written
\bea
\label{winding}
{\bf w} = \sum m_i {\bf w}_i & & m_i \in \mbb{Z},
\eea
then any vector $\bf v$ in the fundamental cell can be uniquely written as 
\bea
{\bf v} = \sum (\alpha_i {\bf v}_i + \beta_i {\bf w}_i) & & \alpha, \beta \in [0,1). 
\eea
As ${\bf v}_i$ and ${\bf w}_i$ are linearly independent, every point is expressible as a linear combination of the 
two sets of vectors.
Because ${\bf v}_i$ are lattice vectors, and ${\bf w}_i$ returns to the brane, all cases with
$\alpha$ or $\beta > 1$ are reducible to the range given. Finally, if there were two distinct expressions
for $\bf v$, then by taking the difference we would find a winding mode not expressible in the form
of (\ref{winding}), in contradiction of the original assumption.

We now define $(W_A)_{ij} = {\bf w}_i \cdot {\bf w }_j$. The winding contribution to the partition function is
\be
\label{windingcont}
\sum_{m_i} \exp \left( -2\pi t m_i (W_A)_{ij} m_j \right) \to \frac{l^{{\dim(W_A)\over2}}}{(\det W_A)^{\half}} \sum_{m_i} 
\exp \left( -\pi l m_i (W_A^{-1})_{ij} m_j \right). 
\ee
Now, for any set of vectors ${\bf u}_i$, $(\det({\bf u}_i \cdot {\bf u}_j))^{\half}$ gives the volume
spanned by that set of vectors. Together, ${\bf v}_i$ and ${\bf w}_i$ span the fundamental cell. As
$$
 (\textrm{\bf Vol. of fund.  cell}) = (\textrm{\bf Vol. along brane}) \times
 (\textrm{\bf Vol. transverse to brane})
$$
we can conclude that
\be
(\det W_A)^{\half} = \frac{(\det g)^{\half}}{(\det M_A)^\half}.
\ee 
Combining (\ref{momcont}) and (\ref{windingcont}), we see that in the $l \to \infty$ limit the 
tree channel lattice factor is simply
\be
\label{normalisation}
\frac{(\det M_A)}{(\det g)^\half}
\ee
which has a nice geometric interpretation and also determines the normalisation of the boundary state 
for branes wrapping toroidal cycles, as that is found by comparison with  the 1-loop annulus amplitude.

There is likewise a nice formula for the intersection number of two branes wrapping distinct 3-cycles. 
Suppose we have
two branes, spanned by vectors ${\bf v}_i, {\bf v'}_i$ such that for each brane (\ref{branewrap}) is
satisfied. We can write each vector in components ${\bf v}_i  = \sum v_{ij} {\bf e}_j$. We then define
\be
\label{intersectionnumber}
I = \det \left( \begin{array}{cccc} 
v_{11} & v_{12} & \ldots & v_{16} \\
v_{21} & v_{22} & \ldots & v_{26} \\
\ldots & \ldots & \ldots & \ldots \\
v'_{31} & v'_{32} & \ldots & v'_{36}  \end{array} \right).
\ee
$I$ measures the number of fundamental cells spanned by the combination of the two branes. 
Now, if the two branes
span $I$ fundamental cells, then it means that they span $I$ points equivalent to the origin. That is, we have $I$
solutions to
\bea
\label{braneinter}
\sum \alpha_i {\bf v}_i + \sum \beta_i {\bf v}'_i \equiv 0 & & \alpha_i, \beta_i \in [0,1).
\eea
However, (\ref{braneinter}) is equivalent to the existence of $I$ sets of $(\alpha_i, \beta_i)$ such that
\bea
\sum \alpha_i {\bf v}_i \equiv - \sum \beta_i {\bf v}'_i & & \alpha_i, \beta_i \in [0,1),
\eea
and therefore $I$ is nothing else than the intersection number of the two branes. Formula (\ref{intersectionnumber})
generalises the familiar case of a 2-torus, where branes with wrapping numbers $(m_1, n_1)$, $(m_2, n_2)$
have intersection number $m_1 n_2 - n_1 m_2$.

The above formulae are simple and require no detailed calculation. In the case of the M\"obius
strip and Klein bottle amplitudes, it would be nice to have an equally elegant way of computing the
lattice contributions and intersection numbers, given the action of $\Omega R$ on the lattice.
Even though $R$ will just act as a projection on the modes, such a formula has eluded us. 
We must therefore explicitly compute modes and intersection points invariant under $\Omega R$.

Suppose we have found generators of momentum modes ${\bf p}_i$ (Klein bottle)
or vectors ${\bf v}_i$ spanning the brane (annulus or M\"obius strip), and also 
winding modes ${\bf w}_i$ in the lattice. Then we define
\bea
(M_{KB})_{ij} & = & {\bf p}_i \cdot {\bf p}_j \nonumber \\
(M_{MS})_{ij} = (M_{A})_{ij} & = & {\bf v}_i \cdot {\bf v}_j \nonumber \\
(W_{KB})_{ij} = (W_{MS})_{ij} & = & {\bf w}_i \cdot {\bf w}_j. 
\eea
Then, if $\dim(M) = \dim(W) = n$, the tree channel lattice mode contributions are
\be
\begin{array}{lclclc}
KB: & \frac{4^n}{(\det M_{KB} \det W_{KB})^{\frac{1}{2}}} & A: & 
\frac{(\det M_A)}{(\det g)^{\half}} & MS: &
\frac{4^n (\det M_{MS})^{\frac{1}{2}}}{(\det W_{MS})^{\frac{1}{2}}}.
\end{array}
\ee
Once we have computed the lattice modes, twisted sector fixed points, and brane
intersection
numbers, we can write down the tree-channel amplitudes as prescribed in
the appendix.

\section{The massless spectrum}
\label{spectrum}

\subsection{The closed string spectrum}

The computation of the closed string spectrum follows the pattern outlined in \cite{bgk99a}.
Closed string states in the $\Theta^k$ twisted sector live at $\Theta^k$ fixed points and must
be invariant under both the orbifold and orientifold projection. It is
simplest first to work out
the orbifold states and then to analyse their behaviour under the action of $\Omega R$. In general,
the fixed points of a $\Theta^k$ twisted sector decompose into orbits of maximal length $k$ 
under the action of $\Theta$. If a fixed point is in an orbit of length $N$, then under $\Theta$
the oscillator part of a state $| L \rangle \otimes | R \rangle$ can
have a phase of $e^{\frac{2 \pi i}{N}}$ or
any multiple thereof.

Once we have the orbifold states, we keep only those invariant under $\Omega R$. For an orbit taken onto itself,
we keep the symmetric part of the NS-NS sector and the antisymmetric part of the R-R sector. For orbits
exchanged among themselves, symmetrisation and anti-symmetrisation results in a single full copy of both sectors
being retained. The net result is some number of chiral
and vector multiplets in each twisted sector. The total number of such multiplets in each sector is given
by the contribution of that sector to $h^{1,1} + h^{2,1}$ (cf \cite{Klein}).

\subsection{The open string spectrum}

In orientifold models the open string sector is determined by tadpole cancellation. This determines
whether the action of $\Omega R$ and $\Theta^{\frac{N}{2}}$ on the Chan-Paton indices is the
symplectic or orthogonal projection. Once $\gamma_{\Omega R}$ and $\gamma_{\frac{N}{2}}$ are known,
to find the spectrum we simply look for states invariant under their action \cite{GimonP}.

The projection is determined by the relative sign of the M\"obius
strip amplitude. In the Klein bottle amplitude
(\ref{kb}), the $\pm$ sign in front of each twisted sector is determined by the action of $\Omega R$ on the ground state.
\be
\Omega R |0 \rangle \otimes |0 \rangle_{NSNS} = \pm |0 \rangle \otimes |0 \rangle_{NSNS}
\ee
This in turn is fixed by the requirement that $\Omega R |p_1 \rangle \otimes |p_2 \rangle_{NSNS} 
=  |p_1 \rangle \otimes |p_2 \rangle_{NSNS}$, where $ |p_1 \rangle$ and $| p_2 \rangle$ are physical NS states
and thus bosonic. Using our conventions (\ref{conventions}), $(\Omega R)
\psi_r \tilde{\psi}_r (\Omega R) ^{-1} = - \psi_r \tilde{\psi}_r$.
Thus, if the ground state is physical we obtain a leading + sign and if the ground state is
unphysical we obtain a leading - sign. The leading sign in the annulus
amplitude is given by
$(-1)^F |0 \rangle$. As $(-1)^F$ determines whether or not a state is
physical through the GSO projection, the result is that in any twisted
sector the annulus and Klein bottle always have the same sign.

The M\"obius strip signs are more delicate. For odd orbifolds (i.e. the $\mbb{Z}_3$ and the $\mbb{Z}_7$) the 
only non-trivial action on the Chan-Paton indices is that of $\Omega R$. Strings stretching between the $6_i$
and $6_{(i + 2k)}$ branes contribute to the $\Theta^k$ twisted sector under an appropriate insertion of 
$\Omega R \Theta^{\lambda}$. For even orbifolds, $(i,i+2k)$ strings contribute to the $\Theta^k$ and $\Theta^{k+\frac{N}{2}}$
sectors.

The leading sign in the M\"obius amplitude, for an insertion of $\Omega R \Theta^{\lambda}$, is given by the action of
$\Omega R \Theta^\lambda$ on the ground state. This sign has several contributions. First,
$\Omega R |0\rangle_R = - |0\rangle_R$, e.g. see
\cite{Dabolkar}. $\Theta^\lambda$ may also have some action 
on the ground state. Finally,
twisted open strings are located at brane intersection points, which may be interchanged under the action of $\Omega R
\Theta^{\lambda}$. Symmetrising and anti-symmetrising these may also give extra signs. Once all the signs are fixed, 
we can then use tadpole cancellation to determine $\gamma_{\Omega R}$ and $\gamma_{\frac{N}{2}}$.

There are several issues arising in the computation of the open string sector. First, in the above procedure we obtain
sector-by-sector tadpole cancellation. This is more stringent than necessary; the vanishing of R-R flux just requires that
the overall tadpole cancel. Secondly, there are cases where tadpole
cancellation does not seem to determine
the form of $\gamma_{\Omega R}$ or $\gamma_{\frac{N}{2}}$. These arise
when the partition function in a twisted sector contains a $\vartheta \left[ \half \atop
  \half \right]$ part, which vanishes. Indeed, for $\mbb{Z}_4$
orbifolds all twisted sector partition functions vanish.

We take the view that these vanishing sectors in a certain sense vanish accidentally. To determine the 
open string spectrum, we require that we still obtain tadpole cancellation under a small formal deformation of the 
twist, e.g. $(\frac{1}{4}, \frac{1}{4}, \half) \to (\frac{1}{4},
\frac{1}{4} + \epsilon, \half + \epsilon)$. This 
procedure allows a determination of the spectrum and,
applied to the cases studied in \cite{bgk99}\cite{bgk99a}, results in
the same spectrum as found there.

\section{Examples of factorisable orientifolds  revisited}
\label{factorisable}
As previously explained, we can relax the requirement that the complete
projector appear in all three (Klein bottle, annulus and M\" obius
strip) amplitudes, yielding models which generically have
different gauge groups carried by $6_{2i}$ and $6_{2i+1}$
branes. In some cases the gauge groups are the same, but the massless
open string spectrum is not invariant under the exchange of the two gauge
group factors, similarly to the $\mathbb{Z}'_6$ case in
\cite{bgk99a}. We found that all the various $\Omega R$ implementations
for the $\mathbb{Z}_4, \mathbb{Z}_6, \mathbb{Z}'_6 $ models yield consistent solutions \footnote{One can also revisit
  the 6-dimensional orientifolds of \cite{bgk99}. We again find consistent
  models, and the computed spectrum is anomaly free.}. Their closed and open string
spectra \footnote{We thank Gabriele Honecker for alerting us to the
presence of vector multiplets in the closed string untwisted sector.}
are shown in tables \ref{466'closed},\ref{4open} and
\ref{66'open}. We use $\bf{1_0}$ to denote a $U(2)$ singlet uncharged
under the gauge group.
\begin{table}[ht]
\caption{Closed string spectra} 
\label{466'closed}
\begin{center}
\begin{tabular}{|l|c|c|c|c|}
\hline
Model & $\Theta^0$ & $\Theta+\Theta^{-1}$  & $\Theta^2 (+\Theta^{-2})$  & $\Theta^3$\\
\hline 
$\mathbb{Z}_4$ ({\bf AAA}) & 5{\mbox C}+1V & 16{\mbox C} & 16{\mbox C} & absent\\
$\mbb{Z}_4$ ({\bf AAB}) & 5{\mbox C}+1V & 12{\mbox C}+4{\mbox V} & 16\mbox{C} &
absent\\
$\mathbb{Z}_6$ ({\bf AAA}) & 4C+1V & 2C+1V & 10C+5V & 10C+1V\\
$\mathbb{Z}_6$ ({\bf BBB}) & 4C+1V & 3C    & 15C    & 10C+1V\\
$\mathbb{Z}'_6$ ({\bf AAA}) & 4C & 7C+5V & 14C+4V & 10C+2V\\
$\mathbb{Z}'_6$ ({\bf BBB}) & 4C & 9C+3V & 18C & 10C+2V\\
\hline 
\end{tabular}
\end{center}
\end{table}
\begin{table}[ht]
\caption{The $\mbb{Z}_4$ open string spectra}
\label{4open}
\begin{center} 
\begin{tabular}{|l|l|c|c|c|}
\hline
Model & $(6_i, 6_i)$  & $(6_i, 6_{i+1})$ & $(6_i, 6_{i+2})$ \\
\hline
{\bf AAA} & $U(16)\times U(4)$ & 1C ${\bf (16, \overline{4})} \oplus {\bf (\overline{16},
  4)}$ &
2C $ {\bf (256,1)}$ +  8C ${\bf (1,16)}$\\
 & (1V+1C) ${\bf (256,1)}\oplus {\bf (1,16)}+$ & & \\
 
& 2C ${\bf(120,1)}\oplus {\bf(\overline{120},1)}+$ & &\\
 & 2C ${\bf (1,6)}\oplus {\bf (1,\overline{6})}$ & &\\
\hline
{\bf AAB} & $U(8)\times U(2)$ & 2C ${\bf (8,\overline{2})} \oplus {\bf (\overline{8},
  2)}$ &
2C ${\bf (64,1_0)}$ + 8C ${\bf (1,4)}$\\
& (1V+1C) ${\bf (64,1_0)}\oplus {\bf (1,4)}+$ & & \\
& 2C ${\bf (28,1_0)}\oplus {\bf (\overline{28}, 1_0)}+$ & &\\
& 2C ${\bf (1,1)} \oplus {\bf (1, \overline{1})}$ & &\\
\hline
\end{tabular}
\end{center}
\end{table}
\begin{center}
\begin{table}[ht]
\caption{$\mathbb{Z}_6, \mathbb{Z}_6^\prime$ open spectra}
\label{66'open}
\begin{center}
\begin{tabular}{|l|l|c|c|c|c|}
\hline
Model & $\left( 6_{i},6_{i} \right)$ &
        $\left( 6_{i},6_{i+1} \right)$  &
        $\left( 6_{i},6_{i+2} \right)$  &
        $\left( 6_{i},6_{i+3} \right)$  \\
\hline
$\mathbb{Z}_6$ & $U(2) \times U(2)$ &
        1C ${\bf (2,\overline{2})} \oplus $ &
        8C ${\bf (3,1_0)} \oplus {\bf (\overline{3},1_0)}+$ &
        4C ${\bf (2,\overline{2})} \oplus$ \\
$\bf{AAA}$ & (1C+1V) ${\bf (4,1_0)} \oplus {\bf (1_0,4)}+$ &
${\bf (\overline{2},2)}$ & 5C ${\bf (4,1_0)}+$ 1C ${\bf (1_0,\overline{4})}$ & ${\bf (\overline{2},2)}$\\
& 2C ${\bf (1,1_0)} \oplus {\bf (\overline{1},1_0)}+$
& & & \\
& 2C ${\bf (1_0,1)} \oplus {\bf (1_0,\overline{1})}$  &  & &\\
\hline
$\mathbb{Z}_6$ & $U(2) \times U(2)$ &
        3C ${\bf (2,\overline{2})} \oplus $ &
        24C ${\bf (3,1_0)} \oplus {\bf (\overline{3},1_0)}+$ &
        4C ${\bf (2,\overline{2})} \oplus$ \\
${\bf BBB}$ & (1C+1V) ${\bf (4,1_0)} \oplus {\bf (1_0,4)}+$ &
${\bf (\overline{2},2)}$ & 15C ${\bf (4,1_0)}$ +  3C ${\bf (1_0,\overline{4})}$ & ${\bf (\overline{2},2)}$\\
& 2C ${\bf (1,1_0)} \oplus {\bf (\overline{1},1_0)}+$
& & & \\
& 2C ${\bf (1_0,1)} \oplus {\bf (1_0,\overline{1})}$ & &  &\\
\hline
${\mathbb{Z}'_6}$ & $U(2) \times U(2)$ &
       2C ${\bf (2,\overline{2})} \oplus $ &
        1C ${\bf (1,1_0)} \oplus {\bf (\overline{1},1_0)}+$ &
        4C ${\bf (2,\overline{2})} \oplus $ \\
{\bf AAA} & (1C+1V) ${\bf (4,1_0)} \oplus {\bf (1_0,4)}+$ &
${\bf (\overline{2},2)}$ & 3C ${\bf (1_0,1)} \oplus {\bf
  (1_0,\overline{1})} +$ & ${\bf (\overline{2},2)}$ \\
& 2C ${\bf (1,1_0)} \oplus {\bf (\overline{1},1_0)}+$
& & 1C ${\bf (4,1_0)}+$ 3C 
        ${\bf (1_0,4)}$ & \\
& 2C ${\bf (1_0,1)} \oplus {\bf (1_0,\overline{1})}$ & &  &\\
\hline
${\mathbb{Z}'_6}$ & $U(2) \times U(2)$ &
        6C ${\bf (2,\overline{2})} \oplus$ &
        9C ${\bf (1,1_0)} \oplus {\bf (\overline{1},1_0)}+$ &
        4C ${\bf (2,\overline{2})} \oplus $ \\
${\bf BBB}$ & (1C+1V) ${\bf (4,1_0)} \oplus {\bf (1_0,4)}+$ &
${\bf (\overline{2},2)}$ & 3C ${\bf (1_0,1)} \oplus {\bf (1_0,\overline{1})} +$ & ${\bf (\overline{2},2)}$\\
& 2C ${\bf (1,1_0)} \oplus {\bf (\overline{1},1_0)}+$
& & 9C ${\bf (4,1_0)}$ +
        3C ${\bf (1_0,4)}$ & \\
& 2C ${\bf (1_0,1)} \oplus {\bf (1_0,\overline{1})}$ & &  &\\
\hline
\end{tabular}
\end{center}
\end{table}
\end{center}
Since for the factorisable $\mathbb{Z}_4$ models with
$v={1\over4}(1,1,-2)$, $\Omega R \Theta$ takes an
$\bf{A}$-type lattice into a $\bf{B}$-type lattice in the first two tori, and 
thus the $\bf{AAA}$ and $\bf{BBA}$, $\bf{AAB}$ and $\bf{BBB}$ to be equivalent to each other. Similarly,
models $\bf{AAA}$ and $\bf{BBA}$, $\bf{AAB}$ and $\bf{BBB}$ for $\mathbb{Z}_6$ as well as
$\bf{AAA}$ and $\bf{BAB}$, $\bf{ABA}$ and $\bf{BBB}$ for $\mathbb{Z}'_6$ are equivalent.

\section{Examples of non-factorisable orientifolds}
\label{Models}

In this section we discuss in some detail a couple of completely new examples of 
$\Omega R$ orientifolds on non-factorisable orbifolds. Employing the general formalism developed
in section 2 and section 3 we consider the solutions to the tadpole
cancellation conditions where we place the D6-branes parallel 
to the orientifold planes. 
The closed string spectrum for all models considered in this section
is computed as in section \ref{spectrum} and appears in table \ref{closedspectrum}.

\begin{table}[ht]
\caption{Closed string spectra}
\label{closedspectrum}
\begin{center} 
\begin{tabular}{|l|c|c|c|c|c|c|c|}
\hline
Model & $\Theta^0$ & 
$\Theta+\Theta^{-1}$  & $\Theta^2 +\Theta^{-2}$  & 
$\Theta^3 +\Theta^{-3} $ &  $\Theta^4 \left(+\Theta^{-4}\right)$ 
&  $\Theta^5 +\Theta^{-5}$ & $\Theta^6$ \\ 
       
\hline 
$\mathbb{Z}_7 ({\bf A})$ & 3{\mbox C} & 4{\mbox C}+3{\mbox V} 
& 4{\mbox C}+3{\mbox V} & 4{\mbox C}+3{\mbox V} & absent & absent  & absent \\
$\mbb{Z}_7 ({\bf B})$ & 3C & 7C & 7C & 7C & absent & absent & absent \\
$\mathbb{Z}_8 ({\bf AA})$ & 4C & 8C & 8C & 8C &10C & absent & absent \\
$\mathbb{Z}_8 ({\bf BA})$ & 4C & 6C+2V & 8C & 6C+2V & 10C & absent & absent \\
$\mathbb{Z}_8^{'} ({\bf AA})$ & 3C & 4C & 10C & 4C & 9C & absent & absent \\
$\mathbb{Z}_8^{'} ({\bf BA})$ & 3{\mbox C} & 4{\mbox C} & 9{\mbox C} + 1{\mbox V} & 4{\mbox C} & 9{\mbox C} & absent & absent \\ 
$\mathbb{Z}_{12} ({\bf AA})$ & 3{\mbox C} & 2{\mbox C}+1{\mbox V} & 2{\mbox C}+1{\mbox V}& 
6{\mbox C} & 6{\mbox C} + 3{\mbox V} & 2{\mbox C} + 1{\mbox V}& 7{\mbox C} \\
$\mathbb{Z}_{12} ({\bf BA})$ & 3{\mbox C} & 3{\mbox C} & 3{\mbox C} & 6{\mbox C} & 9{\mbox C}& 3{\mbox C} & 7{\mbox C}\\
$\mathbb{Z}_{12}^{'} ({\bf AA})$ & 4{\mbox C} & 4{\mbox C} & 2{\mbox C} & 8{\mbox C} & 10{\mbox C} & 4{\mbox C} & 6{\mbox C}\\
$\mathbb{Z}_{12}^{'} ({\bf BA})$ & 4{\mbox C} 
& 3{\mbox C}+1{\mbox V} & 2{\mbox C} & 6{\mbox C} + 2{\mbox V} & 10{\mbox C} & 3{\mbox C} + 1{\mbox V}& 6{\mbox C}\\
\hline 
\end{tabular}
\end{center}
\end{table}

\subsection{$\mbb{Z}_7$ Model : $A_7$ with $v = \frac{1}{7}(1,2,-3)$}
\label{z7section}
\begin{figure}[ht]
\begin{center}
\makebox[10cm]{
\epsfxsize=15cm
\epsfysize=5cm
\epsfbox{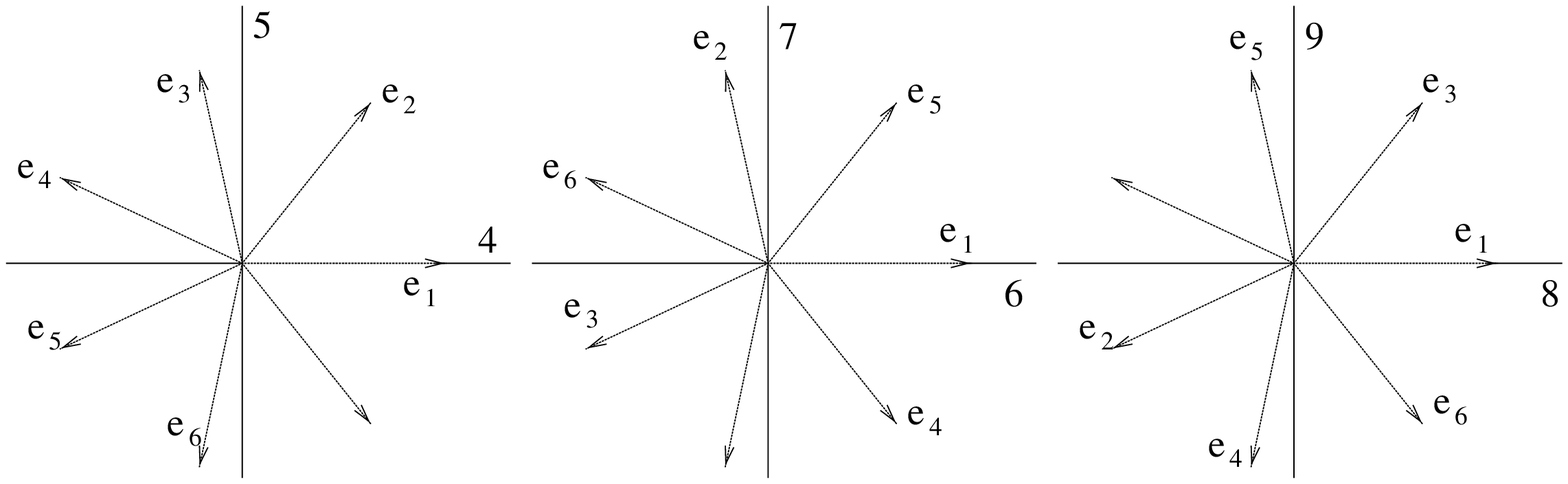}}
\end{center}
\caption{The $\mathbb{Z}_7$ lattice}
\label{z7latticepicture}
\end{figure}

The $SU(7)$ algebra has six root vectors, denoted by ${\bf e}_i$. They are
of equal magnitude (taken to be $1$), and $g_{ij} = {\bf e}_i \cdot {\bf e}_j$ is given by
\begin{equation}
g_{ij} = \left( \begin{array}{cccccc} 1 & -\frac{1}{2} & 0 & 0 & 0 & 0
\\
-\frac{1}{2} & 1 & -\frac{1}{2} & 0 & 0 & 0 \\
0 & -\frac{1}{2} & 1 & -\frac{1}{2} & 0 & 0 \\
0 & 0 & -\frac{1}{2} & 1 & -\frac{1}{2} & 0 \\
0 & 0 & 0 & -\frac{1}{2} & 1 & -\frac{1}{2} \\
0 & 0 & 0 & 0 & -\frac{1}{2} & 1 \end{array} \right)
\end{equation}
The Weyl element $\Gamma_i$ reflects across the plane perpendicular to a root
vector.
Its action on a vector ${\bf x}$  is given by
\begin{equation}
\Gamma_i {\bf x} = {\bf x} - 2\frac{{\bf e}_i \cdot {\bf x}}{{\bf e}_i \cdot {\bf e}_i} {\bf e}_i
\end{equation}
The Coxeter element $\omega = \Gamma_1 \Gamma_2 \ldots \Gamma_6$
acts as
\begin{equation}
\label{z7action}
\begin{array}{cclr} \omega {\bf e}_i & = & {\bf e}_{i+1} &  i =1,2,3,4,5 \\
\omega {\bf e}_6 & = & -{\bf e}_1 -{\bf e}_2 -{\bf e}_3 -{\bf e}_4 -{\bf e}_5 -{\bf e}_6 & \end{array}
\end{equation}
$\omega$ manifestly satisfies $\omega^7 = 1$. As discussed in section
\ref{definitions}, 
the action of $\omega$ can be written as
\bea
Z_i & \to & \exp(2\pi i v_i) Z_i \nonumber \\
\bar{Z}_i & \to & \exp(-2\pi i v_i) \bar{Z}_i, 
\eea
where $v_i = (\frac{1}{7}, \frac{2}{7}, -\frac{3}{7})$ and the $Z_i$ are
coordinates in some orthogonal complex
planes. 
The lattice vectors are visualised as in figure \ref{z7latticepicture}.


%

By inspection, $\Omega R$ acts
crystallographically as
\bea
{\bf e}_1 & \to & {\bf e}_1 \nonumber \\
{\bf e}_2 & \to & -{\bf e}_1 -{\bf e}_2 -{\bf e}_3 -{\bf e}_4 -{\bf e}_5 -{\bf e}_6 \nonumber \\
{\bf e}_3 & \to & {\bf e}_6 \nonumber \\
{\bf e}_4 & \to & {\bf e}_5 \nonumber \\
{\bf e}_5 & \to & {\bf e}_4 \nonumber \\
{\bf e}_6 & \to & {\bf e}_3 .
\eea
In the Klein bottle amplitude, lattice modes only contribute
in the untwisted sector. The momentum
modes are in the dual lattice and have the general form
\be
\alpha(2{\bf e}_1) + \beta (2{\bf e}_3 + 2{\bf e}_6) + \gamma (2{\bf e}_4 + 2{\bf e}_5),
\ee
where $\alpha, \beta, \gamma \in \mathbb{Z}$. The winding modes are
\be 
\label{z7kbwind} 
\alpha'({\bf e}_1 + 2{\bf e}_2 + {\bf e}_3 + {\bf e}_4 + {\bf e}_5 + {\bf e}_6) + \beta'({\bf e}_3 -{\bf e}_6) + \gamma'({\bf e}_4
- {\bf e}_5) \ee
with $\alpha', \beta', \gamma' \in \mathbb{Z}$. We then get
\begin{eqnarray}
M_{KB} & = & 4\left( \begin{array}{ccc} 1 & 0 & 0 \\ 0 & 2 & -1\\ 0 & -1 & 1
\end{array} \right) \\
W_{KB} & = & \left( \begin{array}{ccc} 2 & -1 & 0 \\ -1 & 2 & -1 \\ 0 & -1 & 3
\end{array} \right).
\end{eqnarray}
The untwisted lattice factor is then $\frac{64}{(\det M_{KB})^{\frac{1}{2}}(\det
W_{KB})^{\frac{1}{2}}} = \frac{8}{\sqrt{7}}$.
The $SU(7)$ lattice has 7 fixed points in each twisted sector. These are
$\frac{n}{7}({\bf e}_1 + 2{\bf e}_2 + 3{\bf e}_3 +4{\bf e}_4 + 5{\bf e}_5 + 6{\bf e}_6)$, where $n=0,1,\ldots,6$.
It is easily verified that only
the $n=0$ case is invariant under $\Omega R$.

We can now write down the tree channel Klein bottle amplitude. As the zero point energy
in all sectors is negative, 
$ \Omega R |0 \rangle \otimes |0 \rangle = - |0 \rangle \otimes |0 \rangle$, and all sectors carry a leading minus sign
\begin{eqnarray}
KB & = & c \int_0^\infty 16 l dl \left( \frac{-1}{16l^4}
(\Theta^0)\frac{8l^3}{\sqrt{7}} - \frac{1}{2l}(\Theta) -
\ldots \frac{1}{2l}(\Theta^6) \right) \nonumber \\
& = & c \int_0^\infty dl \frac{8}{\sqrt{7}} \left( -(\Theta^0) -
\sqrt{7}(\Theta^1) - \sqrt{7}(\Theta^2) - \ldots
- \sqrt{7}(\Theta^6) \right).
\label{z7kb}
\end{eqnarray}
As $\sin(\frac{\pi}{7})\sin(\frac{2\pi}{7})\sin(\frac{4\pi}{7}) =
\frac{\sqrt{7}}{8}$,
we actually have the complete projector $\prod_i 2\sin(\pi v_i)$
appearing here.


\begin{figure}[ht]
\begin{center}
\makebox[10cm]{
\epsfxsize=15cm
\epsfysize=5cm
\epsfbox{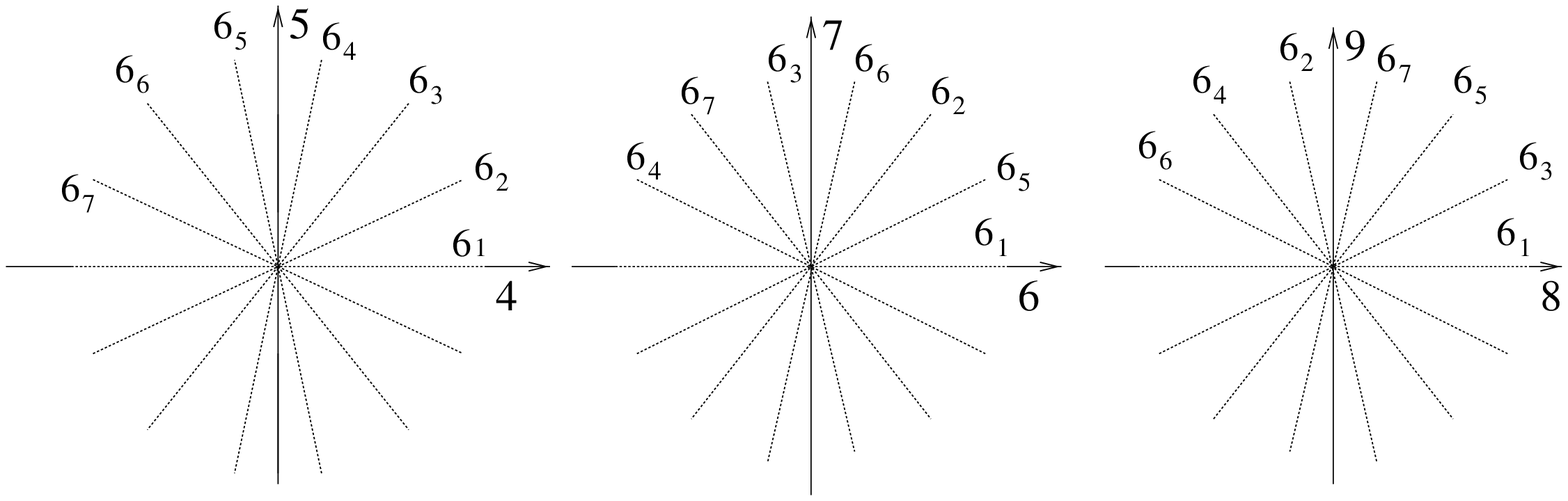}}
\end{center}
\caption{The $\mathbb{Z}_7$ branes}
\label{z7branepicture}
\end{figure}

%
In the limit $l \to \infty$, the amplitude (\ref{z7kb}) diverges.
We add D-branes as shown in figure \ref{z7branepicture}. 
The volume occupied by the $1$ brane is
\be
\alpha' {\bf e}_1 + \beta'({\bf e}_3 + {\bf e}_6) + \gamma'({\bf e}_4 + {\bf e}_5)
\ee
with $\alpha', \beta', \gamma' \in [0,1)$ and therefore
\be
M_A = \left( \begin{array}{ccc} 1 & 0 & 0 \\ 0 & 2 & -1 \\ 0 & -1 & 1
\end{array} \right).
\ee
As explained in section \ref{Lattice}, the lattice mode contribution for the
annulus
is $\frac{(\det M_A)}{(\det g)^\half}
= \frac{8}{\sqrt{7}}$. There are actually no non-trivial brane intersections in the
twisted sectors, and so the annulus amplitude is
\be
A = M^2 c \int l dl \left( \frac{-1}{16l^4} (\Theta^0) \frac{8l^3}{\sqrt{7}} -
\frac{1}{2l}(\Theta) - \ldots -
\frac{1}{2l}(\Theta^6) \right)
\ee
which can be simplified to 
\be
\label{z7a}
A = M^2 c \int \frac{dl}{2 \sqrt{7}} \left( -(\Theta^0) - \sqrt{7} (\Theta) -
\ldots \sqrt{7}(\Theta^6) \right)
\ee
As with the Klein bottle amplitude (\ref{z7kb}), in all sectors the ground state is unphysical
and so $(-1)^F |0 \rangle = - |0 \rangle$.

Finally we need to consider the M\"obius strip amplitude. The twisted sector
intersection numbers remain trivial.
The lattice modes must be invariant under $\Omega R$. As described in section \ref{Lattice}, 
we have
\begin{eqnarray}
M_{MS} = M_A & = & \left( \begin{array}{ccc} 1 & 0 & 0 \\ 0 & 2 & -1 \\ 0 & -1 & 1
\end{array} \right) \\
W_{MS} = W_{KB} & = & \left( \begin{array}{ccc} 2 & -1 & 0 \\ -1 & 2 & -1 \\ 0 & -1 & 3
\end{array} \right)
\end{eqnarray}
with a resulting lattice factor for the M\"obius strip amplitude of $64 \frac{(\det
M_{MS})^{\frac{1}{2}}}{(\det W_{MS})^{\frac{1}{2}}}
= \frac{64}{\sqrt{7}}$.
The M\"obius amplitude reads
\be
MS = +M c \int 16 l dl \left( \frac{1}{2^8 l^4} (\Theta^0) \frac{64l^3}{\sqrt{7}} +
\frac{1}{4l}(\Theta) + \ldots +
\frac{1}{4l}(\Theta^6) \right)
\ee
giving
\begin{equation}
\label{z7ms}
MS = +M c \int dl \frac{4}{\sqrt{7}} \left( (\Theta^0) + \sqrt{7}(\Theta) + \ldots
\sqrt{7}(\Theta^6) \right).
\end{equation}
We have in equation (\ref{z7ms}) fixed $\textrm{tr} (\gamma_{\Omega R \Theta^k}^T \gamma_{\Omega R \Theta^k}^{-1}) = +M$
in all sectors.
Extracting the leading divergences from equations (\ref{z7kb}), (\ref{z7a}) and
(\ref{z7ms}),
the tadpole cancellation conditions are
\begin{equation}
(M^2 -8M + 16) = 0,
\end{equation}
implying $M=4$ and the branes carrying  an $SO(4)$ gauge group.

The closed string spectrum was given in table \ref{closedspectrum}. For the open strings,
the $\Omega R$ projection is the $SO(n)$ projection in all sectors. In each twisted sector, there is one
massless oscillator state located at the origin. 
This being an odd orbifold, there is no additional $\Theta^{\frac{N}{2}}$ projection
to concern us and the open string spectrum is as in table \ref{z7openspectrum}. 

The above action of $R$ is not the only crystallographic implementation. We could also implement $R$ at a rotation of
$\Theta^{\half}$ in the $\bf{B}$ orientation. 
If we repeat the above
analysis, we find that all amplitudes are multiplied by 7. This does not affect the tadpole cancellation condition and we again 
get an $SO(4)$ gauge group. This behaviour is exactly analogous to that encountered for the $\mbb{Z}_3$ orientifold in
\cite{bgk99a}.

\begin{table}[ht]
\caption{The $\mbb{Z}_7$ open string spectrum}
\label{z7openspectrum}
\begin{center} 
\begin{tabular}{|l|c|c|}
\hline
Sector & $\mbb{Z}_7$(\textbf{A}) & $\mbb{Z}_7$(\textbf{B}) \\
\hline
$(6_i, 6_{i})$       & \textrm{1V}$\bf(6)$ + \textrm{3C}($\bf 6$) & \textrm{1V}$\bf(6)$ + \textrm{3C}($\bf 6$) \\ 
$(6_i, 6_{i \pm 1})$ & 1C ($\bf 6$) & 7C ($\bf 6$)\\
$(6_i, 6_{i \pm 2})$ & 1C ($\bf 6$) & 7C ($\bf 6$) \\ 
$(6_i, 6_{i \pm 3})$ & 1C ($\bf 6$) & 7C ($\bf 6$) \\
\hline
\end{tabular}
\end{center}
\end{table}

\subsection{$\mbb{Z}_8^{'}$ Model: $B_2 \times B_4$ with $v = \frac{1}{8}(2,1,-3)$}
\label{z8psection}
\begin{figure}[ht]
\begin{center}
\makebox[10cm]{
\epsfxsize=15cm
\epsfysize=5cm
\epsfbox{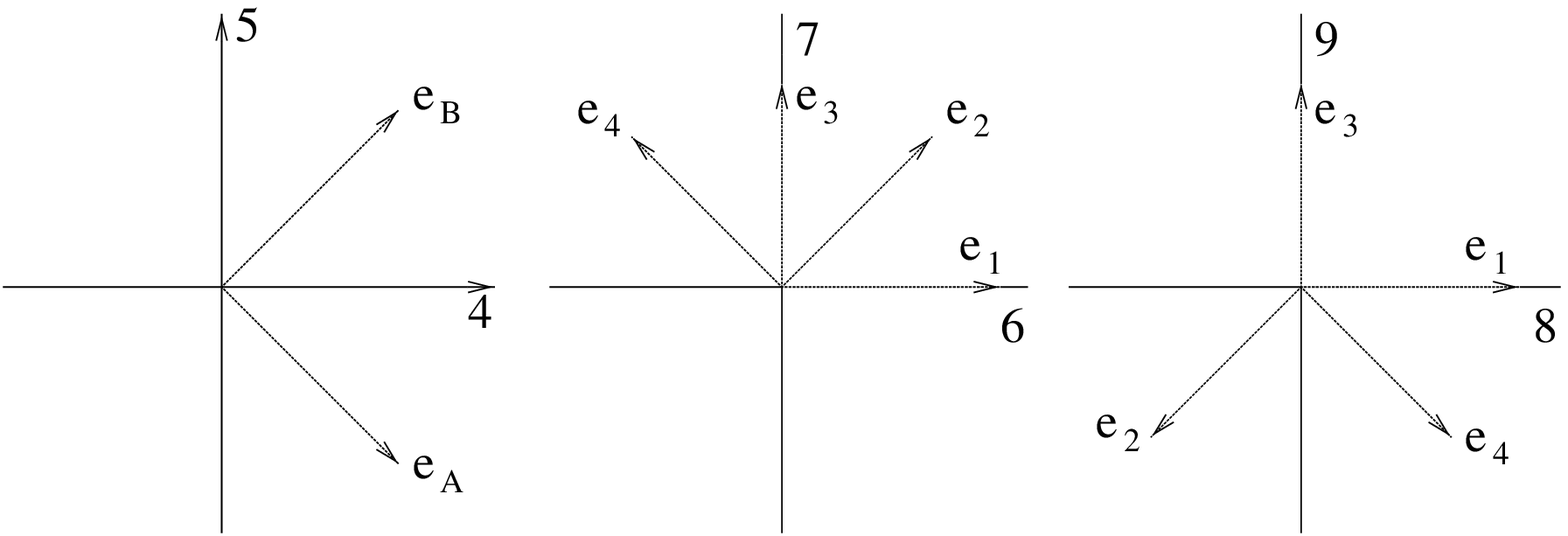}}
\end{center}
\caption{The $\mathbb{Z}_8^{'}$ lattice}
\label{z8latticepicture}
\end{figure}

The $\mbb{Z}_8^{'}$ case has several subtleties typical of even
orientifolds. For $\mbb{Z}_8^{'}$, $\Theta$ has
$v=\frac{1}{8}(2,1,-3)$ and is given by the Coxeter element of $B_2 \times
B_4$. Using ${\bf b}_i$ to denote the simple roots of
$B_4$,
\be
b_{ij} = {\bf b}_i \cdot {\bf b}_j = \left( \begin{array}{cccc}
2 & -1 & 0 & 0 \\
-1 & 2 & -1 & 0 \\
0 & -1 & 2 & -2 \\
0 & 0 & -2 & 4 \end{array} \right).
\ee
We can verify that
\begin{eqnarray}
\omega {\bf b}_1 & = & {\bf b}_2 \nonumber\\
\omega {\bf b}_2 & = & {\bf b}_3 \nonumber \\
\omega {\bf b}_3 & = & {\bf b}_1 + {\bf b}_2 + {\bf b}_3 + {\bf b}_4 \nonumber\\
\omega {\bf b}_4 & = & -2{\bf b}_1 -2{\bf b}_2 -2{\bf b}_3 -{\bf b}_4. 
\end{eqnarray}
The ${\bf b}_i$ are of unequal size and thus not appropriate for the
approach outlined in section \ref{Lattice}.
It is therefore useful to define
\begin{eqnarray}
{\bf e}_1 & = & {\bf b}_1 \nonumber\\
{\bf e}_2 & = & {\bf b}_2 \nonumber  \\
{\bf e}_3 & = & {\bf b}_3 \nonumber\\
{\bf e}_4 & = & {\bf b}_1 + {\bf b}_2 + {\bf b}_3 + {\bf b}_4. 
\end{eqnarray}
The ${\bf e}_i$ are of equal magnitude and by inspection generate the
same lattice as ${\bf b}_i.$ 
satisfy $\omega {\bf e}_i = {\bf e}_{i+1}$ for $i = 1,2,3$
and $\omega {\bf e}_4 = -{\bf e}_1$. 
By inspection, they generate the same lattice as the
${\bf b}_i$. Denoting the lattice vectors of $B_2$ by ${\bf e}_A$,
${\bf e}_B$, and normalising all basis vectors to unity, we then have
\begin{eqnarray}
\omega {\bf e}_A = {\bf e}_B \quad \omega {\bf e}_1 &=& {\bf
  e}_2 \quad \omega {\bf e}_2 = {\bf e}_3 \nonumber\\
\omega {\bf e}_B = -{\bf e}_A \quad \omega {\bf e}_3 &=& {\bf e}_4 \quad
  \omega {\bf e}_4 = -{\bf e}_1.
\end{eqnarray}
\be
\label{z8metric}
g_{ij} = {\bf e}_i \cdot {\bf e}_j = \left( \begin{array}{cccccc}
1 & 0 & 0 & 0 & 0 & 0 \\
0 & 1 & 0 & 0 & 0 & 0 \\
0 & 0 & 1 & -\frac{1}{2} & 0 & \frac{1}{2} \\
0 & 0 & -\frac{1}{2} & 1 & -\frac{1}{2} & 0 \\
0 & 0 & 0 & -\frac{1}{2} & 1 & -\frac{1}{2} \\
0 & 0 & \frac{1}{2} & 0 & -\frac{1}{2} & 1 \end{array} \right).
\ee
The lattice vectors are shown in figure \ref{z8latticepicture}. The action of $R$ is given by
\bea
{\bf e}_A & \leftrightarrow & {\bf e}_B  \nonumber\\
{\bf e}_1 & \leftrightarrow & {\bf e}_1  \nonumber\\
{\bf e}_2 & \leftrightarrow & -{\bf e}_4  \nonumber\\
{\bf e}_3 & \leftrightarrow & -{\bf e}_3. 
\eea


%

There are two independent insertions in
the Klein bottle trace (\ref{kb}), $\Omega R$
and $\Omega R \Theta$. 
Under $\Omega R$, the lattice modes in the untwisted sector are
\bea
\label{z8kbmmor}
{\bf p}_{\Omega R} &=& l({\bf e}_A+{\bf e}_B) + m(2{\bf e}_1) + n({\bf e}_2 - {\bf e}_4), \quad l,m,n \in \mathbb{Z} \nonumber \\
{\bf w}_{\Omega R} &=& l'({\bf e}_A-{\bf e}_B)+m'{\bf e}_3 +n({\bf e}_2 + {\bf e}_4), \quad  l',m',n' \in \mathbb{Z}
\eea
and under $\Omega R \Theta$ we have
\bea
{\bf p}_{\Omega R \Theta} 
&=& l{\bf e}_A + m({\bf e}_1-{\bf e}_4 +{\bf e}_2 -{\bf e}_3) +2n({\bf e}_1-{\bf e}_4), \quad l,m,n \in \mathbb{Z} \nonumber \\
{\bf w}_{\Omega R \Theta} &=& l'{\bf e}_B + m'({\bf e}_1+{\bf e}_4) +n'({\bf e}_2+{\bf e}_3), \quad  l',m',n' \in \mathbb{Z}
\eea
From (\ref{z8kbmmor}) we obtain
\begin{eqnarray}
M_{KB,\Omega R} & = & \left( \begin{array}{ccc} 2& 0 & 0 \\0 & 4 & -2 \\0 & -2
& 2 \end{array} \right) \\
W_{KB, \Omega R} & = & \left( \begin{array}{ccc} 2 & 0 & 0 \\0 & 1 &-1 \\0 & -1
& 2 \end{array} \right).
\end{eqnarray}
The lattice contribution in the untwisted sector for the $\Omega R$
insertion is then
\be
\frac{64}{(\det M_{KB,\Omega R})^\frac{1}{2} (\det W_{KB,\Omega
R})^{\frac{1}{2}}} = 16
\ee
and it is easily verified that the $\Omega R \Theta$ insertion gives the same
factor.

Lattice modes are also present in the $\Theta^4$ twisted sector, which leaves
the $B_2$ lattice
invariant. In this case, the lattice contribution is
\be
\frac{4}{(\det M)^{\frac{1}{2}}(\det W)^{\frac{1}{2}}} = \left\{
\begin{array}{lr} 2 & \Omega R \textrm{ insertion} \\ 4 & \Omega R \Theta \textrm{ insertion.}
\end{array} \right.
\ee
For the twisted sectors, we also need to know the number of fixed points
invariant under $\Omega R$ and $\Omega R\Theta$. In the $\Theta$ twisted sector,
the $B_2$ lattice has by inspection one non-trivial fixed point, $\frac{1}{2}({\bf e}_A + {\bf e}_B)$. In the
$B_4$
lattice, fixed points under $\Theta$ satisfy
\begin{equation}
\omega \left(a_1 {\bf e}_1 + a_2 {\bf e}_2 + a_3 {\bf e}_3 + a_4 {\bf e}_4 \right) \equiv a_1 {\bf e}_1 + a_2 {\bf e}_2
+ a_3 {\bf e}_3 + a_4 {\bf e}_4
\end{equation}
where $a_i \equiv a_i + \mathbb{Z}$.
This is solved
to give $a_i = \frac{1}{2}$ as the one non-trivial solution. 
There are then a total of 4 $\Theta$ fixed points,
having the form
\be
\left( \begin{array}{c} \frac{1}{2} \\ 0 \end{array} \right) ({\bf e}_A + {\bf e}_B)
+ \left( \begin{array}{c} \frac{1}{2} \\ 0 \end{array} \right) ({\bf e}_1 + {\bf e}_2 + {\bf e}_3
+ {\bf e}_4).
\ee
These are all invariant under the action of $\Omega R$ and $\Omega R \Theta$.
In general, we calculate fixed points using our knowledge of the action of
$\omega$, and then
use the action of $\Omega R$ to explicitly check which are invariant. 
The number of fixed
points in the other twisted sectors is shown in table \ref{Z8fixedpoints}. 
As the $\Theta^{N-k}$ sector mirrors the $\Theta^k$ sector in its fixed point structure, we only show 
the first four twisted sectors.

\begin{table}
\caption{Fixed point structure for the $\mbb{Z}_8^{'}$ $B_2 \ti B_4$ orbifold}
\centering
\vspace{3mm}
\label{Z8fixedpoints}
\begin{tabular}{|c|c|c|c|}
\hline
& Fixed points & No. invariant under $R$ & No. invariant under
$R\Theta$ \\
\hline
$\Theta$ & 4 & 4 & 4 \\
$\Theta^2$ & 16 & 8 & 8 \\
$\Theta^3$ & 4 &4 & 4 \\
$\Theta^4$ & 16 & 8 & 4 \\
\hline
\end{tabular}

\end{table}

Combining the lattice modes and the number of invariant fixed points,
we can now evaluate the tree-channel Klein bottle amplitude
\be
-c \int_0^{\infty} 16 dl \left( (\Theta^0) + 2(\Theta) + 4(\Theta^2) + 2(\Theta^3) 
- 4 (\Theta^4) +2(\Theta^5) - 4(\Theta^6) +2(\Theta^7)  \right).
\label{z8kb}
\ee
We observe that equation (\ref{z8kb}) contains the complete
projector. 


\begin{figure}[ht]
\begin{center}
\makebox[10cm]{
\epsfxsize=15cm
\epsfysize=5cm
\epsfbox{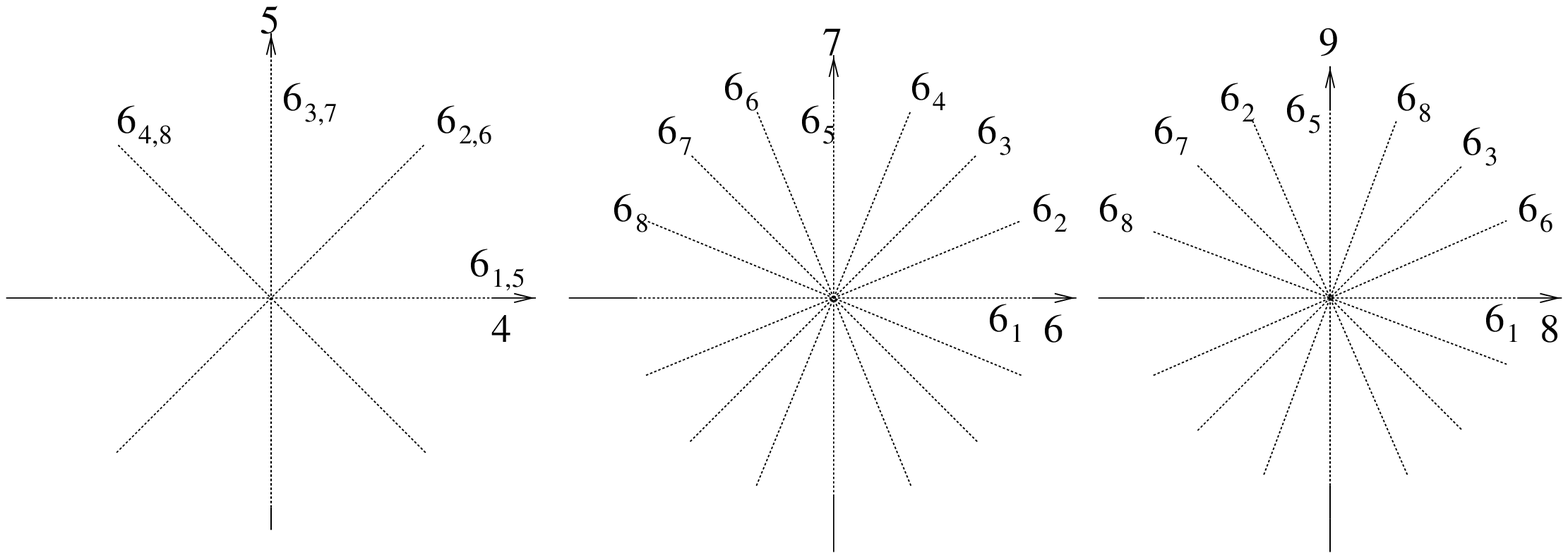}}
\end{center}
\caption{The $\mathbb{Z}_8^{'}$ branes}
\label{z8'branepicture}
\end{figure}

%

D6-branes are then added as in figure \ref{z8'branepicture}, with $M$
branes on each stack.
There are two independent contributions to the
annulus amplitude, coming from strings originating on the $6_{1+2n}$ and $6_{2n}$
branes.
\begin{table}
\caption{Brane intersection points}
\centering
\vspace{3mm}
\label{z8intersections}
\begin{tabular}{|c|c|c|c|c|c|}
\hline
Sector & \# & Location & Sector & \# & Location \\
\hline
(1,2) & 1 &  0 & (2,3) & 1 & 0 \\
(1,3) & 2 & $(0,\frac{1}{2})({\bf e}_A + {\bf e}_B)$ & (2,4) & 2 & $(0,\frac{1}{2})({\bf e}_1 + {\bf e}_2 + {\bf e}_3
+ {\bf e}_4)$\\
(1,4) & 1 & 0 & (2,5) & 1 & 0 \\
(1,5) & 2 & $\frac{1}{2}({\bf e}_2 - {\bf e}_4)$ & (2,6) & 4 & $(0,\frac{1}{2})({\bf e}_2 - {\bf e}_3)
+ (0,\frac{1}{2})({\bf e}_1 - {\bf e}_4)$ \\
\hline
\end{tabular}
\end{table}
The $6_{1}$ brane occupies the volume
\begin{equation}
\alpha({\bf e}_A + {\bf e}_B) + \beta {\bf e}_1 + \gamma({\bf e}_2 - {\bf e}_4)
\end{equation}
so that
\be
M_{A,11} = \left( \begin{array}{ccc} 2 & 0 & 0 \\ 0 & 1 & -1 \\ 0 & -1 & 2
\end{array}
\right).
\ee
The lattice factor in the untwisted sector for ($6_1$, $6_1$) strings is then
$\frac{(\det M_{A,11})}{(\det g)^{\frac{1}{2}}} = 4$. 
The $6_{2}$ brane occupies the volume
\begin{equation}
\alpha' {\bf e}_B + \beta' ({\bf e}_1 + {\bf e}_2) + \gamma' ({\bf e}_3 - {\bf e}_4)
\end{equation}
so that for ($6_2$,$ 6_2$) strings  we likewise obtain a lattice factor of $4$.
In the $\Theta^4$ `twisted' sector, the ($6_1$,$6_5$) and ($6_2$,$6_6$) brane pairs
coincide along the first torus. Here the lattice modes give a factor of 2
for ($6_1,6_5$) strings and 1 for ($6_2$,$6_6$) strings. 

The intersection numbers can be computed as described in section
\ref{Lattice} and are listed in table
\ref{z8intersections} together with their locations. The locations are necessary for the M\"obius strip amplitude and
must be found by explicit computation, e.g. to find ($6_1$, $6_2$) intersection points we look for non-trivial solutions of
\be
\alpha({\bf e}_A + {\bf e}_B) + \beta {\bf e}_1 + \gamma({\bf e}_2 - {\bf e}_4) \equiv
\alpha' {\bf e}_B + \beta' ({\bf e}_1 + {\bf e}_2) + \gamma' ({\bf e}_3 - {\bf e}_4). 
\ee
We can now write down the annulus amplitude
\begin{equation}
-cM^2 \int l dl \left( \frac{1}{16l^4} (\Theta^0)4 +
\frac{1}{2l}(\Theta)
+\frac{1}{2l}2(\Theta^2) + \frac{1}{2l}(\Theta^3) +
\frac{l}{4l^2} 4
(\Theta^4) + \ldots\right)
\end{equation}
which can be brought to the form
\begin{equation}
A = -{M^2 c} \int \frac{dl}{4} \left( (\Theta^0) + 2(\Theta) + 4(\Theta^2) +
2(\Theta^3) - 4(\Theta^4) + \ldots \right).
\end{equation}
This has the same form as the Klein bottle amplitude (\ref{z8kb}).

For the M\"obius amplitude, as discussed in section \ref{Lattice}, we find 
\be
W_{MS,11} = W_{KB,\Omega R} = \left( \begin{array}{ccc} 2 & 0 & 0 \\ 0 & 1 & -1 \\ 0 & -1 & 2
\end{array}
\right)
\ee
\be
M_{MS,11} = M_{A,11} = \left( \begin{array}{ccc} 2 & 0 & 0 \\ 0 & 1 & -1 \\ 0 & -1 & 2
\end{array}
\right).
\ee
The lattice factor for ($6_1$, $6_1$) strings is then
$64 \frac{(\det M_{MS,11})^{\frac{1}{2}}}{(\det W_{MS,11})^{\frac{1}{2}}} = 64$ and a
similar treatment for ($6_2$, $6_2$) strings yields 64, as well.

No other lattice modes contribute to the M\"obius amplitude. The $6_1$ and $6_5$ branes
have a coincident
direction and so could in principle have lattice modes. However, ($6_1$, $6_5$) strings
are invariant under
an insertion of $\Omega R \Theta^2$ in the trace. As this acts as a reflection
on the coincident
direction, there are no invariant lattice modes. Finally, all intersection points are invariant
under the appropriate insertion of $\Omega R \Theta^k$ except those in the ($6_2$, $6_6$) sector.
Here, only two of the four points are invariant.

We then obtain the following tree channel M\"obius strip amplitude
\begin{equation}
\label{z8ms}
Mc \int 4dl \left( (\Theta^0) + 2(\Theta) + 4(\Theta^2) +
2(\Theta^3) - 4(\Theta^4) + \ldots \right).
\end{equation}
To obtain (\ref{z8ms}) we have fixed the action of the various Chan-Paton matrices; this determines
the open string spectrum.
Extracting the leading divergence from the three amplitudes,
we get
\begin{equation}
(M-8)^2 = 0
\end{equation}
implying that we need stacks of 8 D-branes to cancel tadpoles and that the
gauge group is $U(4) \times U(4)$. 

The above represents a \textbf{BA}
$\Omega R$ implementation. 
We can also consider the  $\bf{AA}$ $\Omega R$
orientation. The tree channel Klein bottle amplitude is then
\bea
K &=& -c \int_0^{\infty} dl \biggl( (16+4) (\Theta^0) + 8(2+2) (\Theta) + 8(8+2) (\Theta^2) + 8(2+2) (\Theta^3)
- \nonumber \\ 
&& 8(8+2) (\Theta^4) + \ldots \biggr)
\label{z8aakb}
\eea
Supposing that there are $M$ $6_{2k}$ and $N$ $6_{2k+1}$ branes, the tree channel annulus amplitude
takes the form
\bea
A &=& -c \int \frac{dl}{32} \biggl( (2M^2+8N^2) (\Theta^0) + 16MN (\Theta) + 8(M^2+4N^2)(\Theta^2) \nonumber \\ 
&&+ 16MN (\Theta^3) - 8(M^2+4N^2) (\Theta^4) + \ldots \biggr)
\eea
and the M\" obius amplitude is similarly seen to be
\bea
\label{z8aams}
M &=& c \int 4dl \Biggl( {1\over2}(M+N) (\Theta^0) + {1\over2}(M+4N) (\Theta) + 2(M+N) (\Theta^2) + \nonumber\\
&&   {1\over2}(M+4N) (\Theta^3) - 2 (M+N)(\Theta^4) + \ldots \Biggr).
\eea
The tadpole cancellation conditions imply $M=16, N=4.$ This means that the gauge group is
$U(8)\times U(2).$ The open string spectra for the two orientations
are shown in Table \ref{z8'openspectra}.

\begin{table}[ht]
\caption{The $\mbb{Z}_8^{'}$ open string spectra}
\label{z8'openspectra}
\begin{center} 
\begin{tabular}{|l|c|c|}
\hline
Sector & $\mbb{Z}_8^{'}$(\bf{AA}) & $\mbb{Z}_8^{'}$(\bf{BA}) \\
\hline
$(6_i, 6_{i})$ & \textrm{(1V + 1C)   }$(\bf 64, 1) \oplus (\bf 1, 4)$+  & \textrm{(1V + 1C)   }$(\bf 16, 1) \oplus (\bf 1, 16)$+ \\
& \textrm{2C }$(\bf 28,1_0) \oplus (\overline{28},1_0) \oplus (1,1) \oplus (1, \overline{1})$  & 
\textrm{2C }$(\bf 6,1_0) \oplus (\overline{6},1_0) \oplus (1,6) \oplus (1, \overline{6})$ \\
\hline
$(6_i, 6_{i \pm 1})$ & 1C $\bf (8, \overline{2}) \oplus (\overline{8}, 2)$ & 1C $\bf (4, \overline{4}) \oplus (\overline{4}, 4)$\\
\hline
$(6_i, 6_{i \pm 2})$ & \textrm{1C} $\bf (28,1_0) \oplus
(\overline{28},1_0)$ + \textrm{4C }$\bf (1,1) \oplus (1, \overline{1})$ & 
\textrm{2C }$\bf (6,1) \oplus (\overline{6},1) \oplus (1,6)
\oplus (1,\overline{6})$\\ 
\hline
$(6_i, 6_{i \pm 3})$ & 1C $\bf (8, \overline{2}) \oplus (\overline{8}, 2)$& 1C $\bf (4, \overline{4}) \oplus (\overline{4}, 4)$\\
\hline
$(6_i, 6_{i + 4})$ & 2C $\bf (64,1_0)$ + 4C $\bf (1,4)$  
& 2C $\bf (16,1)$ + 4C $\bf (1, 16)$ \\
\hline
\end{tabular}
\end{center}
\end{table}

\subsection{$\mbb{Z}_8$ Model: $D_2 \times B_4$ with $v = \frac{1}{8}(-4,1,3)$}
\label{z8section} 
The $\mbb{Z}_8$ lattice is very similar to the $\mbb{Z}_8^{'}$
lattice, the only difference being the replacement of $B_2$ by
$D_2$. Indeed, the lattice picture is a simple modification of
\ref{z8latticepicture}.
The action of $\omega$ on the lattice basis ${\bf e}_i$ is
\begin{eqnarray}
\omega {\bf e}_A = -{\bf e}_A \quad \omega {\bf e}_1 &=& {\bf
  e}_2 \quad \omega {\bf e}_2 = {\bf e}_3 \nonumber\\
\omega {\bf e}_B = -{\bf e}_B \quad \omega {\bf e}_3 &=& {\bf e}_4 \quad
  \omega {\bf e}_4 = -{\bf e}_1.
\end{eqnarray}
Here ${\bf e}_A, {\bf e}_B$
are a basis for the $D_2$ and the ${\bf e}_i, i=1, \ldots, 4$ are
a basis for $B_4$, as in section \ref{z8psection}. The lattice metric
for this basis is given by (\ref{z8metric}).
In the same manner as in \ref{z8psection} we obtain consistent models for the two inequivalent $\Omega R$ implementations. 
The calculations are very similar to the $\mathbb{Z}'_8$
orientifold. Gauge groups turn out to be $U(8)\times U(4)$ ({\bf{AA}} case) and $U(4)\times U(2)$
({\bf{BA}} case). Results for open spectra are shown in Table \ref{z8spectra}.

\begin{table}[ht]
\caption{The $\mbb{Z}_8$ open string spectra}
\label{z8spectra}
\begin{center} 
\begin{tabular}{|l|c|c|}
\hline
Sector & $\mbb{Z}_8$(\bf{AA}) & $\mbb{Z}_8$(\bf{BA}) \\
\hline
$(6_i, 6_{i})$ & \textrm{(1V + 1C)   }$(\bf 64, 1) \oplus (\bf 1, 16)$+  & \textrm{(1V + 1C)   }$(\bf 16, 1) \oplus (\bf 1, 4)$+ \\
& \textrm{2C }$(\bf 28,1) \oplus (\overline{28},1) \oplus (1,6) \oplus (1, \overline{6})$  & 
\textrm{2C }$(\bf 6,1_0) \oplus (\overline{6},1_0) \oplus (1_0,1) \oplus (1_0, \overline{1})$ \\
\hline
\rule{0pt}{2.5ex} 
$(6_i, 6_{i \pm 1})$ & 1C $\bf (8, \overline{4}) \oplus (\overline{8}, 4)$ & 2C $\bf (4, \overline{2}) \oplus (\overline{4}, 2)$\\
\hline
\rule{0pt}{2.5ex}
$(6_i, 6_{i \pm 2})$ & \textrm{2C} $\bf (64,1)$ + \textrm{4C }$\bf (1,16)$ & 
\textrm{2C }$\bf (16,1)$ +  \textrm{4C } $\bf (1,4)$ \\ 
\hline
\rule{0pt}{2.5ex}
$(6_i, 6_{i \pm 3})$ & 2C $\bf (8, \overline{4}) \oplus (\overline{8}, 4)$& 4C $\bf (4, \overline{2}) \oplus (\overline{4}, 2)$\\
\hline
$(6_i, 6_{i + 4})$ & 2C $\bf (28,1) \oplus (\overline{28},1)$ + 4C $\bf (1,6) \oplus (1,\overline{6})$  
& 2C $\bf (6,1) \oplus (\overline{6},1)$ + 4C $\bf (1_0 ,1) \oplus
(1_0 ,\overline{1})$ \\
\hline
\end{tabular}
\end{center}
\end{table}

\subsection{$\mbb{Z}_{12}$ Model: $A_2 \times F_4$ with $v = \frac{1}{12}(4,1,-5)$}
\label{z12section}

\begin{figure}[ht]
\begin{center}
\makebox[10cm]{
\epsfxsize=15cm
\epsfysize=5cm
\epsfbox{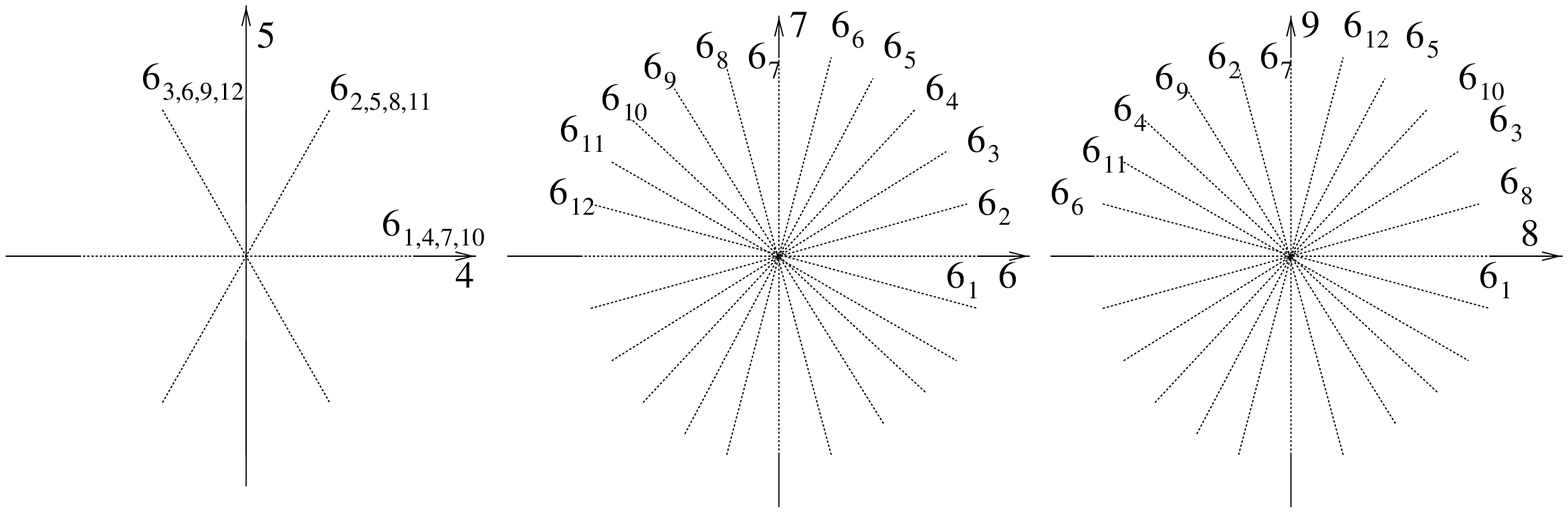}}
\end{center}
\caption{The $\mathbb{Z}_{12}$(\bf{AA}) branes}
\label{z12branepicture}
\end{figure}

%

Here the lattice vectors are visualised as in figure \ref{z12latticepicture}, and the branes are placed
as in figure \ref{z12branepicture}. 
The action of $\omega$ on the lattice basis ${\bf e}_i$ is given by
\bea
\omega {\bf e}_1 = {\bf e}_2\qquad
\omega {\bf e}_2 &=& -({\bf e}_1+{\bf e}_2)\qquad
\omega {\bf e}_3 = {\bf e}_4 \nonumber \\
\omega {\bf e}_4 = {\bf e}_5\qquad
\omega {\bf e}_5 &=& {\bf e}_6\qquad\qquad\qquad
\omega {\bf e}_6 = {\bf e}_5 - {\bf e}_3.
\eea
and also
\be
\label{z12metric}
g_{ij} = {\bf e}_i \cdot {\bf e}_j = \left( \begin{array}{cccccc}
1 & -\frac{1}{2} & 0 & 0 & 0 & 0 \\
-\frac{1}{2} & 1 & 0 & 0 & 0 & 0 \\
0 & 0 & 1 & -\frac{1}{2} & \frac{1}{2} & 0 \\
0 & 0 & -\frac{1}{2} & 1 & -\frac{1}{2} & \frac{1}{2} \\
0 & 0 & \frac{1}{2} & -\frac{1}{2} & 1 & -\frac{1}{2} \\
0 & 0 & 0 & \frac{1}{2} & -\frac{1}{2} & 1 \end{array} \right)
\ee
The calculation follows exactly the same pattern as outlined above for the
$\mbb{Z}_8^{'}$ case, and we find the tadpole cancellation condition
\be
\label{z12tadpole}
(M-4)^2 = 0
\ee
which implies a $U(2) \times U(2)$ gauge group. In this case there are
two consistent implementations of $\Omega R$.
If we rotate the $A_2$ into a \textbf{B} type lattice, then the effect
is to multiply all the amplitudes by 3 which does not modify
the tadpole cancellation conditions. We denote this model by $\mbb{Z}_{12}(\bf{BA})$. The spectra for the 
two models are shown in table \ref{z12openspectrum}. 


\begin{table}[ht]
\caption{The $\mbb{Z}_{12}$ open string spectra}
\label{z12openspectrum}
\begin{center} 
\begin{tabular}{|l|c|c|}
\hline
Sector & $\mbb{Z}_{12}$(\textbf{AA}) & $\mbb{Z}_{12}$(\textbf{BA}) \\
\hline
$(6_i, 6_{i})$ & \textrm{(1V + 1C)   }$(\bf 4, 1) \oplus (\bf 1, 4)$+  & \textrm{(1V + 1C)   }$(\bf 4, 1) \oplus (\bf 1, 4)$+ \\
& \textrm{2C }$(\bf 1,1_0) \oplus (\overline{1},1_0) \oplus (1_0,1) \oplus (1_0, \overline{1})$  & 
\textrm{2C }$(\bf 1,1_0) \oplus (\overline{1},1_0) \oplus (1_0,1) \oplus (1_0, \overline{1})$ \\
\hline
\rule{0pt}{2.5ex} 
$(6_i, 6_{i \pm 1})$ & 1C $\bf (2, \overline{2}) \oplus (\overline{2}, 2)$ & 3C $\bf (2, \overline{2}) \oplus (\overline{2}, 2)$\\
\hline
\rule{0pt}{2.5ex}
$(6_i, 6_{i \pm 2})$ & \textrm{1C }$(\bf 1,1_0) \oplus (\overline{1},1_0) \oplus (1_0,1) \oplus (1_0, \overline{1})$ 
& \textrm{3C }$(\bf 1,1_0) \oplus (\overline{1},1_0) \oplus (1_0,1) \oplus (1_0, \overline{1})$ \\ 
\hline
\rule{0pt}{2.5ex}
$(6_i, 6_{i \pm 3})$ & 2C $\bf (2, \overline{2}) \oplus (\overline{2}, 2)$& 6C $\bf (2, \overline{2}) \oplus (\overline{2}, 2)$\\
\hline
$(6_i, 6_{i \pm 4})$ & 2C $\bf (4,1) \oplus (1,4)$ + & 6C $\bf (4,1) \oplus (1,4)$ + \\
& 1C $\bf (3,1) \oplus (\overline{3},1) \oplus (1,3) \oplus (1,\overline{3})$& 
3C $\bf (3,1) \oplus (\overline{3},1) \oplus (1,3) \oplus (1,\overline{3})$ \\
\hline
\rule{0pt}{2.5ex}
$(6_i, 6_{i \pm 5})$ & 1C $\bf (2, \overline{2}) \oplus (\overline{2}, 2)$ & 3C $\bf (2, \overline{2}) \oplus (\overline{2}, 2)$ \\
\hline
\rule{0pt}{2.5ex}
$(6_i, 6_{i +6})$ & 3C $(\bf 1,1_0) \oplus (\overline{1},1_0) \oplus (1_0,1) \oplus (1_0, \overline{1})$ &
9C $(\bf 1,1_0) \oplus (\overline{1},1_0) \oplus (1_0,1) \oplus (1_0, \overline{1})$ \\
& 1C $\bf (3,1) \oplus (\overline{3},1) \oplus (1,3) \oplus (1,\overline{3})$ & 
3C $\bf (3,1) \oplus (\overline{3},1) \oplus (1,3) \oplus (1,\overline{3})$ \\
\hline
\end{tabular}
\end{center}
\end{table}

%

We note that the different zero point energies of the twisted sectors result in the $\bf A + \bar{A}$, $\bf S + \bar{S}$, and
\textbf{Adj} representations appearing in the various sectors.

\subsection{$\mbb{Z}_{12}^{'}$ Model: $D_2 \times F_4$ with $v = \frac{1}{12}(-6,1,5)$}
\label{z12psection}
The $\mbb{Z}_{12}^{'}$ case is very similar to the $\mbb{Z}_{12}$ case, the difference being the 
replacement of the $A_2$ by $D_2$. 
The lattice vectors and brane positions are a simple modification of
figures \ref{z12latticepicture} and \ref{z12branepicture}. The action
of $\omega$ on the basis vectors ${\bf e}_i$ for the lattice is:
\bea
\omega {\bf e}_1 = -{\bf e}_1\qquad
\omega {\bf e}_2 &=& -{\bf e}_2\qquad
\omega {\bf e}_3 = {\bf e}_4 \nonumber \\
\omega {\bf e}_4 = {\bf e}_5\qquad
\omega {\bf e}_5 &=& {\bf e}_6\qquad
\omega {\bf e}_6 = {\bf e}_5 - {\bf e}_3
\eea
and the metric on the lattice is
\be
\label{z12pmetric}
g_{ij} = {\bf e}_i \cdot {\bf e}_j = \left( \begin{array}{cccccc}
1 & 0 & 0 & 0 & 0 & 0 \\
0 & 1 & 0 & 0 & 0 & 0 \\
0 & 0 & 1 & -\frac{1}{2} & \frac{1}{2} & 0 \\
0 & 0 & -\frac{1}{2} & 1 & -\frac{1}{2} & \frac{1}{2} \\
0 & 0 & \frac{1}{2} & -\frac{1}{2} & 1 & -\frac{1}{2} \\
0 & 0 & 0 & \frac{1}{2} & -\frac{1}{2} & 1 \end{array} \right).
\ee
Again, there are two consistent implementations of $\Omega R$, arising from reorienting the $D_2$ lattice. This time the
tadpole cancellation conditions are altered, the $\bf AA$ type lattice giving
\be
\label{z12atadpole}
(M-8)^2 = 0
\ee
and the $\bf BA$ type lattice giving 
\be
\label{z12btadpole}
(M-4)^2 = 0.
\ee
The spectra for the two models are shown in table \ref{z12'openspectrum}.


\begin{table}[ht]
\caption{The $\mbb{Z}_{12}^{'}$ open string spectra}
\label{z12'openspectrum}
\begin{center} 
\begin{tabular}{|l|c|c|}
\hline
Sector & $\mbb{Z}_{12}^{'}$(\textbf{AA}) & $\mbb{Z}_{12}^{'}$(\textbf{BA}) \\
\hline
$(6_i, 6_{i})$ & \textrm{(1V + 1C)   }$(\bf 16, 1) \oplus (\bf 1, 16)$+  & \textrm{(1V + 1C)   }$(\bf 4, 1) \oplus (\bf 1, 4)$+ \\
& \textrm{2C }$(\bf 6,1) \oplus (\overline{6},1) \oplus (1,6) \oplus (1, \overline{6})$  & 
\textrm{2C }$(\bf 1,1_0) \oplus (\overline{1},1_0) \oplus (1_0,1) \oplus (1_0, \overline{1})$ \\
\hline
\rule{0pt}{2.5ex} 
$(6_i, 6_{i \pm 1})$ & 1C $\bf (4, \overline{4}) \oplus (\overline{4}, 4)$ & 2C $\bf (2, \overline{2}) \oplus (\overline{2}, 2)$\\
\hline
\rule{0pt}{2.5ex}
$(6_i, 6_{i \pm 2})$ & \textrm{2C }$(\bf 16,1) \oplus (1,16)$ 
& \textrm{2C }$\bf (4,1) \oplus (1,4)$ \\ 
\hline
\rule{0pt}{2.5ex}
$(6_i, 6_{i \pm 3})$ & 2C $\bf (4, \overline{4}) \oplus (\overline{4}, 4)$& 4C $\bf (2, \overline{2}) \oplus (\overline{2}, 2)$\\
\hline
$(6_i, 6_{i \pm 4})$ & 2C $\bf (16,1) \oplus (1,16)$ + & 2C $\bf (4,1) \oplus (1,4)$ + \\
& 4C $\bf (6,1) \oplus (\overline{6},1) \oplus (1,6) \oplus (1,\overline{6})$ & 
\textrm{4C }$(\bf 1,1_0) \oplus (\overline{1},1_0) \oplus (1_0,1) \oplus (1_0, \overline{1})$ \\
\hline
\rule{0pt}{2.5ex}
$(6_i, 6_{i \pm 5})$ & 1C $\bf (4, \overline{4}) \oplus (\overline{4}, 4)$ & 2C $\bf (2, \overline{2}) \oplus (\overline{2}, 2)$ \\
\hline
\rule{0pt}{2.5ex}
$(6_i, 6_{i +6})$ & 4C $\bf (16,1) \oplus (1,16)$ &
4C $\bf (4,1) \oplus (1,4)$ \\
\hline
\end{tabular}
\end{center}
\end{table}

%

\subsection{$\mbb{Z}_{4}$ Model: $A_3 \times A_3$ with $v = \frac{1}{4}(1,-2,1)$}
\label{a3a3section}
In table \ref{lattices}, many of the orbifolds have non-trivial fixed tori in some
twisted sectors. These are contained inside one of the constituent
lattices. For example, the $\mbb{Z}_{12}$ model on the $E_6$ lattice
has a non-trivial fixed torus in the $\Theta^4$ twisted sector. 
In such twisted sectors, we encounter many novel features. 
For example, the closed string winding modes may now be
fractional multiples of lattice vectors. Nonetheless, after all issues
are taken into account we still obtain perfect sector-by-sector
tadpole cancellation.
We will illustrate this by
studying in detail the $\mbb{Z}_4$ orientifold on the $A_3 \times A_3$ lattice.
For compactness, we will only explicitly perform the calculations
in the $\Theta^2$ sector where there is a non-trivial fixed torus. The
calculations in the other sectors are analogous to those studied above
and neither present nor result in problems.


\begin{figure}[ht]
\begin{center}
\makebox[10cm]{
\epsfxsize=15cm
\epsfysize=5cm
\epsfbox{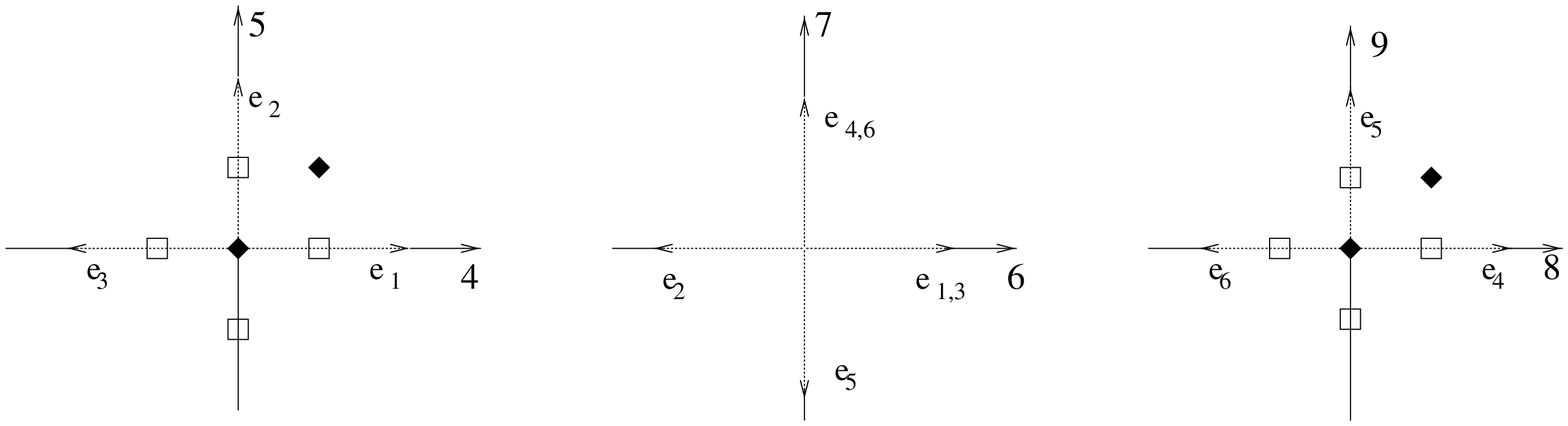}}
\end{center}
\caption{The $A_3 \times A_3$ $\mathbb{Z}_4$ vectors. The black dots
  represent the locations of $\Theta^2$ fixed tori and the squares the
  loci for half-integral winding modes.}
\label{a3a3vectors}
\end{figure}

%
We can draw the $A_3 \ti A_3$ lattice as in figure
\ref{a3a3vectors}. The action of $\omega$ is

\bea
\omega {\bf e}_1 = {\bf e}_2 \qquad\omega {\bf e}_2 & = & {\bf e}_3\qquad
\omega {\bf e}_3 = -{\bf e}_1 - {\bf e}_2 - {\bf e}_3 \nonumber \\
\omega {\bf e}_4 = {\bf e}_5 \qquad\omega {\bf e}_5 & = & {\bf e}_6\qquad
\omega {\bf e}_6 = -{\bf e}_4 -{\bf e}_5 -{\bf e}_6 
\eea
and the metric is given by
\be g_{ij} = {\bf e}_i \cdot {\bf e}_j = \left( \begin{array}{cccccc}
1 & -\half & 0 & 0 & 0 & 0 \\
-\half & 1 & -\half & 0 & 0 & 0 \\
0 & -\half & 1 & 0 & 0 & 0 \\
0 & 0 & 0 & 1 & -\frac{1}{2} & 0 \\
0 & 0 & 0 & -\frac{1}{2} & 1 & -\frac{1}{2} \\
0 & 0 & 0 & 0 & -\frac{1}{2} & 1 \end{array} \right).
\ee
There are three independent implementations of $R$: \textbf{AAA}, \textbf{BAA} and
\textbf{AAB}. Here \textbf{A} and \textbf{B} denote the orientation of the $R$ fixed plane in
each rotation plane; \textbf{A} corresponds to aligning the fixed
plane along the $x$-axis and \textbf{B} at
an angle of $\frac{\pi}{4}$ to this.  In case \textbf{AAA} $R$ acts as 
\bea
{\bf e}_1 & \leftrightarrow & {\bf e}_1 \nonumber \\
{\bf e}_2 & \leftrightarrow & -{\bf e}_1 -{\bf e}_2 -{\bf e}_3 \nonumber \\ 
{\bf e}_3 & \leftrightarrow & {\bf e}_3 \nonumber \\
{\bf e}_4 & \leftrightarrow & -{\bf e}_6 \nonumber \\
{\bf e}_5 & \leftrightarrow & -{\bf e}_5  
\eea
and in case $\bf{BAA}$ the action is  
\bea
{\bf e}_1 & \leftrightarrow & {\bf e}_1 + {\bf e}_2 + {\bf e}_3 \nonumber \\ 
{\bf e}_2 & \leftrightarrow & -{\bf e}_3 \nonumber \\
{\bf e}_4 & \leftrightarrow & -{\bf e}_6 \nonumber \\
{\bf e}_5 & \leftrightarrow & -{\bf e}_5  
\eea
and finally in case $\bf{AAB}$ $R$ acts as
\bea
{\bf e}_1 & \leftrightarrow & {\bf e}_1 \nonumber \\
{\bf e}_2 & \leftrightarrow & -{\bf e}_1 -{\bf e}_2 -{\bf e}_3 \nonumber \\
{\bf e}_3 & \leftrightarrow & {\bf e}_3 \nonumber \\
{\bf e}_4 & \leftrightarrow & {\bf e}_5 \nonumber \\
{\bf e}_6 & \leftrightarrow & -{\bf e}_4 -{\bf e}_5 -{\bf e}_6. 
\eea
We will only explicitly consider the $\bf{AAA}$ $\Omega R$
orientation, as the behaviour for $\bf{BAA}$ and $\bf{AAB}$ is similar.
For the Klein bottle amplitude, we need to know the structure of fixed
points and fixed tori. The $\Theta^2$
sector has 4 fixed tori
\be
\frac{n'}{2}({\bf e}_1 + {\bf e}_2) + \frac{m'}{2}({\bf e}_4 + {\bf e}_5) + \alpha({\bf e}_1 + {\bf e}_3) +
\beta({\bf e}_4 + {\bf e}_6)
\ee
where $n',m' \in \{0,1\}$. From
figure \ref{a3a3vectors}, or by explicit computation, one can see that the action of
$R$ on the fixed tori in general involves a shift as well as a
reflection. As the lattice is non-factorisable, the different rotation 
planes do not decouple from each other. 

In the $\Theta^2$ twisted sector, naively the momentum and
winding modes would be, for an $\Omega R$ insertion in the trace,
\bea
{\bf p} & = & \frac{n}{2}({\bf e}_1 + {\bf e}_3) \nonumber \\
{\bf w} & = & m({\bf e}_4 + {\bf e}_6)
\eea
with $m,n \in \mbb{Z}$, and for an $\Omega R \Theta$ insertion in the
trace,
\bea
{\bf p} & = & \frac{n}{2}({\bf e}_4 + {\bf e}_6) \nonumber  \\
{\bf w} & = & m({\bf e}_1 + {\bf e}_3).
\eea
However, there are two subtle issues that render this invalid. First,
since $R$ acts as
\bea
R: \frac{1}{2}({\bf e}_1 + {\bf e}_2) & \to & \frac{1}{2}({\bf e}_1 + {\bf e}_2) + \frac{1}{2}({\bf e}_1 + {\bf e}_3) \nonumber \\
R: \frac{1}{2}({\bf e}_4 + {\bf e}_5) & \to & \frac{1}{2}({\bf e}_4 + {\bf e}_5) + \frac{1}{2}({\bf e}_4 + {\bf e}_6),
\eea
on the fixed tori labelled by $(n',m')$ $R$ acts as a shift by $\frac{n'}{2}({\bf e}_1 + {\bf e}_3) + 
\frac{m'}{2}({\bf e}_4 + {\bf e}_6)$. Recall that the effect of a translation 
$T:{\bf x} \to {\bf x}+{\bf a}$ on a momentum mode $| {\bf p} \rangle $ is
\be
\label{transphase}
T |{\bf p} \rangle = \exp (2 \pi i {\bf p} \cdot {\bf a}) | {\bf p} \rangle .
\ee 
Therefore, for tori where $n'=1$, momentum modes having
${\bf p} = (1 + 2k)({\bf e}_1 + {\bf e}_3)$ pick up a phase factor 
of $e^{i \pi}$ under the action of $R$. When the sum over momentum modes is 
performed, such modes have no net contribution, because they cancel against similar modes from tori where no
such shift occurs. This results in an effective doubling of the momentum modes. Schematically,
\be
\sum_n (-1)^n \exp (-\pi t n^2 p^2) + \sum_n \exp (-\pi t n^2 p^2) = 2 \sum_n \exp(-4 \pi t n^2 p^2).
\ee
The shift along the winding direction does not have any effect on the
sum in the partition function. Moreover, as 
\bea
R \Theta: \frac{1}{2}({\bf e}_1 + {\bf e}_2) & \to & \frac{1}{2}({\bf e}_1 + {\bf e}_2) \nonumber \\
R \Theta: \frac{1}{2}({\bf e}_4 + {\bf e}_5) & \to & \frac{1}{2}({\bf e}_4 + {\bf e}_5)
\eea
the insertion of $\Omega R \Theta$ has no effect on the
contributing momentum modes.

There is also a subtlety at work for the winding modes. The condition
that a string state be an acceptable orbifold state in the $\Theta^2$
twisted sector is that
\be
X^\mu (\sigma + 2\pi, \tau) = \Theta^2 X^\mu(\sigma, \tau).
\ee
Consider the points marked with a square in figure
\ref{a3a3vectors}. These have the property that, when acted on with
$\Theta^2$, they are brought back to themselves with a shift by either $\half
({\bf e}_4 + {\bf e}_6)$ or $\half ({\bf e}_1 + {\bf e}_3)$. Explicitly, we have for example
\be
\Theta^2 \frac{1}{4}({\bf e}_4 - {\bf e}_6) = \frac{1}{4}({\bf e}_4 - {\bf e}_6) + \half ({\bf e}_4
+ {\bf e}_6)
\ee
It is this shift that allows the existence of winding modes that
are half-integral multiples of lattice vectors. Thus the winding state
\be
X(\sigma, \tau) = \frac{1}{4}({\bf e}_4 - {\bf e}_6) +
\frac{\sigma}{2\pi} \frac{({\bf e}_4 + {\bf e}_6)}{2} + (\textrm{$\tau$ dependence})
\ee
is a legitimate orbifold state in the $\Theta^2$ twisted sector.

Next, we must consider whether these states are invariant under the
action of $\Omega R$ and $\Omega R \Theta$. Under the insertion of
$\Omega R \Theta$, all loci for half-integral winding modes are
exchanged among themselves and there is no net contribution to the
partition function. However, under the insertion of $\Omega R$, four
such loci survive, and we have additional winding modes of the form
\be
X(\sigma, \tau) = \left( \begin{array}{c} \half \\ 0 \end{array} \right) ({\bf e}_1 + {\bf e}_2) \pm \frac{{\bf e}_4 - {\bf e}_6}{4} 
+ \frac{\sigma}{2\pi} \frac{{\bf e}_4 + {\bf e}_6}{2} + (\textrm{$\tau$ dependence}).
\ee
The multiplicity is exactly the same as that of the number of fixed
tori and the net effect is then that the effective winding mode for
the $\Omega R$ insertion is reduced from $({\bf e}_4 + {\bf e}_6)$ to $\half({\bf e}_4 +
{\bf e}_6)$. For the $\Omega R \Theta$ insertion, the net winding mode is
unchanged.

The actual momentum and winding modes appearing in the partition function are
then, for an $\Omega R$ insertion in the trace
\bea
{\bf p} & = & n({\bf e}_1 + {\bf e}_3) \nonumber \\
{\bf w} & = & \frac{m}{2}({\bf e}_4 + {\bf e}_6)
\eea
with $m,n \in \mbb{Z}$, and for an $\Omega R \Theta$ insertion in the
trace,
\bea
{\bf p} & = & \frac{n}{2}({\bf e}_4 + {\bf e}_6) \nonumber \\
{\bf w} & = & m({\bf e}_1 + {\bf e}_3).
\eea

Performing a similar analysis for the other cases, we obtain the following Klein bottle amplitudes
\bea
\textrm{Case \textbf{AAA}} 
& -c \int_0^{\infty} 16 dl \left( (\Theta^0) + 4(\Theta) - 4(\Theta^2) - 4(\Theta^3) \right) &
\label{a3a3kb1} \nonumber \\
& & \nonumber \\
\textrm{Case \textbf{BAA}} & -c \int_0^{\infty} 32 dl \left( (\Theta^0) + 4(\Theta) - 4(\Theta^2) - 4(\Theta^3) \right) &
\label{a3a3kb2} \nonumber \\
& & \nonumber \\
\textrm{Case \textbf{AAB}} & -c \int_0^{\infty} 8 dl \left( (\Theta^0) + 4(\Theta) - 4(\Theta^2) - 4(\Theta^3) \right). &
\label{a3a3kb3}
\eea

Let us now consider the annulus amplitudes. We describe case $\bf{AAA}$ in detail
and quote the results for the other cases. For case $\bf{AAA}$, the branes are at
\bea
6_1 & & \alpha {\bf e}_1 + \beta {\bf e}_3 + \gamma ({\bf e}_4 - {\bf e}_6) \nonumber \\ 
6_2 & & \alpha ({\bf e}_1 + {\bf e}_2) + \beta({\bf e}_4 + {\bf e}_5) + \gamma({\bf e}_4 + {\bf e}_6) \nonumber \\
6_3 & & \alpha {\bf e}_2 + \beta({\bf e}_1 + {\bf e}_3) + \gamma (2{\bf e}_5 + {\bf e}_4 + {\bf e}_6) \nonumber \\
6_4 & & \alpha ({\bf e}_2 + {\bf e}_3) + \beta ({\bf e}_5 + {\bf e}_6) + \gamma ({\bf e}_4 + {\bf e}_6).
\eea
As before, we focus on the sector arising from $(6_1, 6_3)$ and $(6_2,
6_4)$ strings. In each case there are two intersection points: 0 and
$\half({\bf e}_4 + {\bf e}_6)$ for $(6_1, 6_3)$ strings and  0 and $\half({\bf e}_5 +
{\bf e}_6)$ for $(6_2, 6_4)$ strings. For the former case, the lattice modes
are given by
\bea
{\bf p} & = & \frac{n}{2}({\bf e}_1 + {\bf e}_3) \nonumber \\
{\bf w} & = & \frac{m}{2}({\bf e}_4 + {\bf e}_6) 
\eea
and for $(6_2, 6_4)$ strings by 
\bea
{\bf p} & = & \frac{n}{2}({\bf e}_4 + {\bf e}_6) \nonumber \\
{\bf w} & = & \frac{m}{2}({\bf e}_1 + {\bf e}_3).
\eea
Interestingly, the winding modes in the second case are a
composite result of strings starting at two distinct points.
The cases of $m$ even arise from strings starting at ${\bf 0}$; the cases of
$m$ odd arise from strings starting at $\half ({\bf e}_1 + {\bf e}_2)$. Similar
behaviour occurs for the other two cases.
Computing  the annulus amplitude, we obtain
\bea
\textrm{Case \textbf{AAA} } & -cM^2 \int_0^{\infty} \frac{dl}{4} \left( (\Theta^0) + 4(\Theta) - 4(\Theta^2) - 4(\Theta^3) \right) &
\label{a3a3a1} \nonumber \\
& & \nonumber \\
\textrm{Case \textbf{BAA} } & -cM^2 \int_0^{\infty} \frac{dl}{2} \left( (\Theta^0) + 4(\Theta) - 4(\Theta^2) - 4(\Theta^3) \right) &
\label{a3a3a2} \nonumber \\
& & \nonumber \\
\textrm{Case \textbf{AAB} } & -cM^2 \int_0^{\infty} \frac{dl}{8} \left( (\Theta^0) + 4(\Theta) - 4(\Theta^2) - 4(\Theta^3) \right). &
\label{a3a3a3}
\eea

In the computation of the M\"obius strip amplitude, the $\Theta^2$ sector arises from the insertion of $\Omega R \Theta^2$
into the trace, for $(6_1, 6_1)$ and $(6_2, 6_2)$ strings. This insertion leaves one plane invariant, which can then contribute
lattice modes. The contributing modes are then, for case \textbf{AAA},
\bea
(6_1, 6_1) & {\bf p} = n({\bf e}_1 + {\bf e}_3), & {\bf w} = \frac{m}{2}({\bf e}_4 + {\bf e}_6) \nonumber \\
(6_2, 6_2) & {\bf p} = \frac{n}{2}({\bf e}_4 + {\bf e}_6), & {\bf w} = m({\bf e}_1 + {\bf e}_3). 
\eea
Here, new behaviour is seen for the winding modes in the $(6_1, 6_1)$
sector. For the factorisable models studied previously, half-integral
winding modes did not appear in the M\"obius strip amplitude. Instead, a
winding mode doubling was observed. To see where the difference lies, consider
the $(6_1, 6_1)$ winding mode starting at $\bf{0}$ and ending at
$\half({\bf e}_4 - {\bf e}_6)$ and having winding
vector $\half({\bf e}_4 + {\bf e}_6)$. Under $\Omega R \Theta^2$, this is taken to
a winding mode which starts at $\half({\bf e}_4 - {\bf e}_6)$ and ends at $\bf{0}$,
with the winding vector remaining invariant. So, the effect of 
$\Omega R \Theta^2$ on these modes is as a translation,
\be
T: {\bf x} \to {\bf x} + \half({\bf e}_4 - {\bf e}_6).
\ee
As in (\ref{transphase}), this causes momentum modes to pick up a phase
$\exp (\pi i \bf{p} \cdot ({\bf e}_4 - {\bf e}_6))$. For the case above, the momentum
modes are orthogonal to the translation direction and the phase
factor is unity. Therefore the half-integral modes contribute to the
partition function. In the factorisable case, the momentum modes lay
along the translation direction. For half-integral winding modes, the
sum of the momentum modes in the partition function was then 
\be
\sum_n (-1)^n \exp (-\pi t n^2 p^2) 
\ee
which does not give rise to any divergence in the $t\to 0$ limit. 
This can be seen straightforwardly using the Poisson resummation formula.
Having taken the above points into account, we obtain the
following M\"obius strip amplitudes
\bea
\textrm{Case \textbf{AAA} } 
& cM \int_0^{\infty} 4 dl \left( (\Theta^0) + 4(\Theta) - 4(\Theta^2) - 4(\Theta^3) \right) &
\label{a3a3ms1}  \nonumber\\
& & \nonumber \\
\textrm{Case \textbf{BAA} } & cM \int_0^{\infty} 8 dl  \left( (\Theta^0) + 4(\Theta) - 4(\Theta^2) - 4(\Theta^3) \right) &
\label{a3a3ms2}  \nonumber \\
& & \nonumber \\
\textrm{Case \textbf{AAB} } & cM \int_0^{\infty} 2 dl  \left( (\Theta^0) + 4(\Theta) - 4(\Theta^2) - 4(\Theta^3) \right). &
\label{a3a3ms3}
\eea

Comparing the respective Klein bottle, annulus and M\"obius strip
amplitudes, we see that in all three cases the amplitudes factorise
perfectly and we obtain a tadpole cancellation condition
\be
(M - 8)^2 = 0 
\ee
giving a gauge group of $U(4) \times U(4)$. The behaviour encountered
above in the $A_3 \times A_3$ is typical of those models which have a
non-trivial fixed torus in one of the twisted sectors. After taking
into account the subtleties described above, and if necessary using
different numbers of branes on different O-planes, one can cancel
all tadpoles sector-by-sector. Despite the many subtleties involved in
the one-loop calculation, at the end one obtains a simple
answer. Indeed, for the $A_3 \ti A_3$ case above the amplitudes
actually take the form of the complete projector. 

As well as the models written up in detail above, we have also studied
all other cases in table \ref{lattices}. In table \ref{finalresults} we
give tadpole-cancelling solutions for all $\Omega R$ orientations for
the models in table \ref{lattices}. For cases 1,2,5 and 7, some
results have already appeared in \cite{bgk99a}. The spectra for these
models can be determined exactly as for the other cases 
using the methods described in section \ref{spectrum}. 
However, as they would not present any new features we have not
written them out.

\begin{table}
\caption{IIA Orientifolds of $\mbb{Z}_N$ orbifolds with D-branes on O-planes}
\label{finalresults}
\centering
\vspace{3mm}

\begin{tabular}{|cr|c|c|c|c|}
\hline
\multicolumn{2}{|c|}{ Case}
& Lie algebra root lattice & $\Omega R$ orientation & No. of branes &
Gauge Group \\
\hline
\hline
1 & $\mbb{Z}_3$ & $A_2 \times A_2 \times A_2$ & 4 described in \cite{bgk99a} & $M=4$  &
$SO(4)$\\
\hline
2 & $\mbb{Z}_4$ & $A_1 \ti A_1 \ti B_2 \ti B_2$ & \bf{AAA} & $M=32$,
$N=8$ & $U(16) \ti U(4)$\\
& & & \bf{AAB} & $M=16, N=4$ & $U(8) \ti U(2)$\\
& & & \bf{ABA} & $M=16, N=16$ & $U(8) \ti U(8)$\\
& & & \bf{ABB} & $M=8, N=8$ & $U(4) \ti U(4)$\\
\hline
3 & $\mbb{Z}_4$ & $A_1 \ti A_3 \ti B_2$ & \bf{AAA} & $M$=16, $N$=4 &
$U(8) \times U(2)$ \\
& & & \bf{ABA} & $M$ = 16, $N$ = 8 & $U(8) \times U(4)$\\
& & & \bf{AAB} & $M$ = 8, $N$ = 8 & $U(4) \times U(4)$ \\
& & & \bf{ABB} & $M$ = 8, $N$ = 16 & $U(4) \times U(8)$\\
\hline
4 & $\mbb{Z}_4$ & $A_3 \ti A_3$ & \bf{AAA} & $M=N=8$ & $U(4) \times
U(4)$\\
& & & \bf{BAA} & $M$ = $N$ = 8 & $U(4) \times U(4)$\\
& & & \bf{AAB} & $M$ = $N$ = 8 & $U(4) \times U(4)$\\
\hline
5 & $\mbb{Z}_6$ & $A_2 \ti G_2 \ti G_2$ & \bf{AAA} & $M=N=4$ & $U(2)
\ti U(2)$\\
& & & \bf{BBB} & $M=N=4$ & $U(2) \ti U(2)$\\
& & & \bf{ABA} & $M=N=4$ & $U(2) \ti U(2)$\\
& & & \bf{ABB} & $M=N=4$ & $U(2) \ti U(2)$\\
\hline
6 & $\mbb{Z}_6$ & $G_2 \ti A_2 \ti A_2$ & \bf{AA} & $M=N=4$ & $U(2)
\ti U(2)$\\
& & & \bf{BA} & $M$ = $N$ = 4 & $U(2) \times U(2)$\\
\hline
7 & $\mbb{Z}_6'$ & $A_1 \ti A_1 \ti A_2 \ti G_2$ & \bf{AAA} & $M=N=4$ &
$U(2) \ti U(2)$\\
& & & \bf{AAB} & $M=N=4$ & $U(2) \ti U(2)$\\
& & & \bf{BBB} & $M=N=4$ & $U(2) \ti U(2)$\\
& & & \bf{ABB} & $M=N=4$ & $U(2) \ti U(2)$\\
\hline
9 & $\mbb{Z}_6'$ & $A_1 \ti A_1 \ti A_2 \ti A_2$ & \bf{AA} & $M=N=8$ &
$U(4) \ti U(4)$\\
& & & \bf{BA} & $M$ = $N$ = 4 & $U(2) \times U(2)$\\
\hline
10 & $\mbb{Z}_6'$ & $A_1 \ti A_5$ & \bf{AAA} & $M=8$, $N=4$ & $U(4)
\ti U(2)$\\
& & & \bf{AAB} & $M$ = $N$ = 8 & $U(4) \times U(4)$\\
\hline
11 & $\mbb{Z}_7$ & $A_6$ & \bf{A} & $M=4$ & $SO(4)$ \\
& & & \bf{B} & $M = 4$ & $SO(4)$\\
\hline
12 & $\mbb{Z}_8$ & $B_4 \times D_2$ & \bf{AA} & $M=16$,$N=8$ & $U(8)
\ti U(4)$ \\
& & & \bf{BA} & $M = 8$, $N=4$ & $U(4) \times U(2)$\\
\hline
13 & $\mbb{Z}_8$ & $A_1 \times D_5$ & \bf{AA} & $M=N=8$ & $U(4) \ti U(4)$ \\
& & & \bf{AB} & $M = N = 8$ & $U(4) \times U(4)$\\
\hline
14 & $\mbb{Z}_8'$ & $B_2 \ti B_4$ & \bf{AA} & $M=16$, $N=4$ & $U(8)
\ti U(2)$ \\
& & & \bf{BA} & $M= N = 8$ & $U(4) \times U(4)$\\
\hline
15 & $\mbb{Z}_8'$ & $A_3 \ti A_3$ & \bf{A} & $M=8$, $N=4$ & $U(4) \ti
U(2)$\\
\hline
16 & $\mbb{Z}_{12}$ & $A_2 \ti F_4$ & \bf{AA} & $M=N=4$ & $U(2) \ti
U(2)$ \\
& & & \bf{BA} & $M = N = 4$ & $U(2) \times U(2)$\\
\hline
17 & $\mbb{Z}_{12}$ & $E_6$ & \bf{A} & $M=N=4$ & $U(2) \ti U(2)$ \\
\hline
18 & $\mbb{Z}_{12}'$ & $D_2 \ti F_4$ & \bf{AA}  & $M=N=4$ & $U(2) \ti
U(2)$ \\
& & & \bf{BA} & $M = N = 8$ & $U(2) \times U(2)$\\
\hline
\end{tabular}

\end{table}

\section{Conclusions}

In this paper we have developed techniques to construct supersymmetric Type IIA
orientifolds on six-dimensional orbifolds which do not factorise into
three two-dimensional tori. Using these
techniques we have constructed many  new explicit orientifold models, where we
placed the D6-branes parallel to the orientifold planes. 
For some of these models, in particular those containing  non-trivial fixed tori, 
we have encountered some new technical features  in the amplitudes, which 
nicely played together to finally yield consistent solutions to the tadpole
cancellation conditions. All these supersymmetric  Type IIA  orientifold
models are expected to lift up to M-theory compactifications on
singular compact $G_2$ manifolds \cite{km01,aw01,wit01}.

The next step is to move beyond these simple solutions to the tadpole cancellation
conditions and  to study more general (supersymmetric) intersecting D-brane
models in these backgrounds. 
It is expected that these more general intersecting  D6-branes 
give rise to chirality and might be interesting for building  
semi-realistic models. Once interesting models are found, employing
the results of the effective low energy action as described in 
\cite{cim03,ls03,cp03a,ao03,ao03a,kn03,lmrs04,cim04},
the discussion of the phenomenological implications would be the natural 
next step to perform.

\vspace{1cm}

\noindent {\bf Acknowledgements} \\ 

RB is supported by PPARC and JC is grateful to EPSRC for a research studentship. KS thanks Trinity College,
Cambridge for financial support. We are grateful to Gabriele Honecker
and Tassilo Ott for discussions. It is a pleasure to thank Fernando
Quevedo for helpful comments and encouragement. 

\begin{appendix}

\section{Oscillator Formulae}

Here we give the oscillator contribution to the tree channel amplitude for a complex plane twisted by $\theta$. For the closed
string traces, this means that points related by a rotation of $2\pi \theta$ are identified. For the 
open string sectors, this refers to strings stretching between branes at a relative angle $\pi \theta$.
\begin{eqnarray}
\textrm{Klein bottle} & \theta = 0 & \frac{1}{2l} \frac{ \vartheta \left[ \frac{1}{2} \atop 0 \right]}{\eta^3}(2il) \nonumber \\
& |\theta| \in \left(0, \frac{1}{2}\right] & \frac{ \vartheta \left[ \frac{1}{2} \atop |\theta| \right] }
                                       { \vartheta \left[ \frac{1}{2} \atop |\theta| - \frac{1}{2} \right] }(2il) \nonumber\\
\textrm{Annulus} & \theta = 0 & \frac{1}{2l} \frac{ \vartheta \left[ \frac{1}{2} \atop 0 \right]}{\eta^3}(2il) \nonumber\\
& |\theta| \in \left(0, \frac{1}{2}\right] & \frac{ \vartheta \left[ \frac{1}{2} \atop |\theta| \right] }
                                       { \vartheta \left[ \frac{1}{2} \atop |\theta| - \frac{1}{2} \right] } (2il) \\
\textrm{M\"obius strip} & \theta = 0 & 
\frac{1}{4l} \frac{ \vartheta \left[ \half \atop 0 \right] \vartheta \left[ 0 \atop \half \right] }
                  {\eta^3 \vartheta \left[ 0 \atop 0 \right] } (4il) \nonumber\\
& |\theta| \in \left(0, \frac{1}{2} \right] & \left(
\frac{ \vartheta \left[ \half \atop \frac{|\theta|}{2} \right] \vartheta \left[ \half \atop -\frac{|\theta|}{2} \right] 
\vartheta \left[ 0 \atop \frac{|\theta|}{2} - \half \right] \vartheta 
\left[ 0 \atop \half -\frac{|\theta|}{2} \right]}
     { \vartheta \left[ 0 \atop \frac{|\theta|}{2} \right] \vartheta 
\left[ 0 \atop -\frac{|\theta|}{2} \right] \vartheta \left[ \half \atop \frac{|\theta|}{2} - \half \right] 
\vartheta \left[ \half \atop \half -\frac{|\theta|}{2} \right]} (4il) \right)^{\half} \nonumber\\
\textrm{M\"obius strip (2)} & \theta = 0  & 2 \frac{ \vartheta \left[ \half \atop \half \right] \vartheta \left[ 0 \atop 0 \right]}
{ \vartheta \left[ \half \atop 0 \right] \vartheta \left[ 0 \atop \half \right] } (4il) \nonumber\\
& |\theta| \in \left(0, \half \right] & \left( 
\frac{ \vartheta \left[ 0 \atop \frac{|\theta|}{2} \right] \vartheta \left[ 0 \atop -\frac{|\theta|}{2} \right] 
\vartheta \left[ \half \atop \frac{|\theta|}{2} - \half \right] \vartheta 
\left[ \half \atop \half -\frac{|\theta|}{2} \right]}
     { \vartheta \left[ \half \atop \frac{|\theta|}{2} \right] \vartheta 
\left[ \half \atop -\frac{|\theta|}{2} \right] \vartheta \left[ 0 \atop \frac{|\theta|}{2} - \half \right] 
\vartheta \left[ 0 \atop \half -\frac{|\theta|}{2} \right]} (4il) \right)^{\half} \nonumber
\end{eqnarray}

For $(6_i, 6_{i+2k})$ strings, the two M\"obius strip amplitudes correspond to the schematic insertions of
$\Omega R \Theta^k$ and $\Omega R \Theta^{k + \frac{N}{2}}$. For odd orbifolds, only the former case is present.
The former contributes to the $\Theta^k$ twisted sector; the latter to the $\Theta^{k + \frac{N}{2}}$ twisted sector.

The definitions of the $\vartheta$ functions can be found in the appendix of \cite{bgk99a}.
All the amplitudes were transformed to tree channel using the modular transformation properties
\bea
\vartheta \left[ \alpha \atop \beta \right] \left( t^{-1} \right) & = &  
\sqrt{t}\ e^{2\pi i \alpha \beta} 
\vartheta \left[ -\beta \atop \alpha \right] \left( t \right)  \nonumber \\
\eta \left( t^{-1} \right) & = & \sqrt{t}\ \eta ( t ).
\eea

Explicitly writing out the oscillator contribution to the twisted amplitudes is neither interesting nor
enlightening. We have throughout this paper used $(\Theta^ k)$ to denote the appropriate combination of
$\vartheta$ functions for the $\Theta^k$ twisted sector. In general this consists of a product
of the above terms, one for each complex rotation plane. Where applicable, this also includes the sum of lattice modes
$\sum_{n_i} \exp (-\pi l n_i (A)_{ij} n_j)$ that goes to 1 in the $l \to \infty$ limit. The leading numerical factors we 
always show explicitly. As an example, in the case of the $\mbb{Z}_7$ annulus amplitude, $(\Theta^1)$ refers to the $\Theta$
twisted sector, and is given by
\be
(\Theta^1) = \frac{ \vartheta \left[ \half \atop 0 \right] \vartheta \left[ \half \atop \frac{1}{7} \right]
\vartheta \left[ \half \atop \frac{2}{7} \right] \vartheta \left[ \half \atop \frac{3}{7} \right] }
{\eta^3 \vartheta \left[ \half \atop \frac{1}{7} - \half \right] 
\vartheta \left[ \half \atop \frac{2}{7} - \half \right] \vartheta \left[ \half \atop \frac{3}{7} - \half \right]}(2il).
\ee

\section{Lattices}
In this appendix we describe the various lattices in terms of the action of $\Theta$ on the basis
vectors for the lattice on the three planes where $\Theta$ acts as a
rotation. In every case,
the picture describing the components of the basis vectors in each of the three planes, as well
as the metric for that basis and the explicit action of $\Theta$ on the basis are given.
\begin{enumerate}
\item
$\underline{\bf \mbb{Z}_3, A_2\times A_2\times A_2}$

This is discussed in detail in \cite{bgk99a}.
\item
$\underline{\bf \mbb{Z}_4, A_1\times A_1\times B_2\times B_2}$

Discussed in \cite{bgk99a}.
\item
$\underline{\bf \mbb{Z}_4, A_1\times A_3\times B_2}$
\begin{figure}[ht]
\begin{center}
\makebox[10cm]{
\epsfxsize=15cm
\epsfysize=5cm
\epsfbox{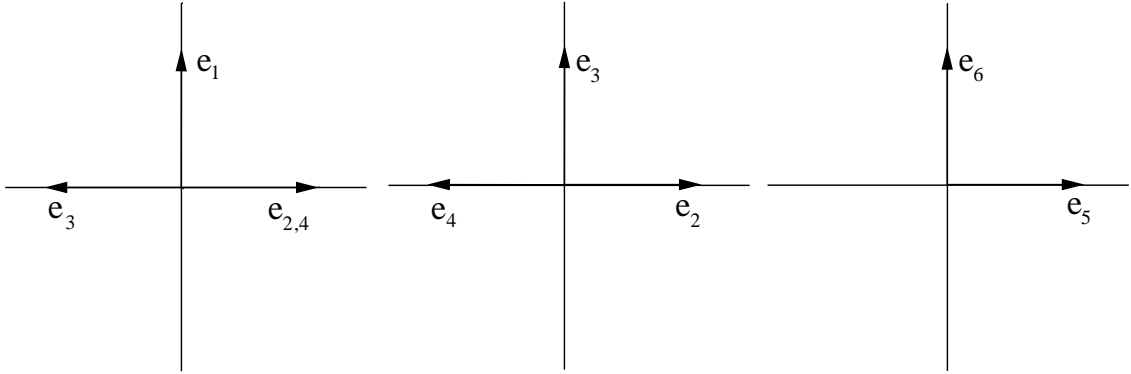}}
\end{center}
\caption{The $\mbb{Z}_4, A_1\times A_3\times B_2$ lattice}
\label{a1a3b2pic}
\end{figure}
%
\begin{eqnarray*}
\omega {\bf e}_1 = -{\bf e}_1\qquad
\omega {\bf e}_2 &=& {\bf e}_3\qquad
\omega {\bf e}_3 = {\bf e}_4\\
\omega {\bf e}_4 = -{\bf e}_2 -{\bf e}_3 - {\bf e}_4\qquad
\omega {\bf e}_5 &=& {\bf e}_6\qquad
\omega {\bf e}_6 = -{\bf e}_5.
\end{eqnarray*}
\be g_{ij} = {\bf e}_i \cdot {\bf e}_j = \left( \begin{array}{cccccc}
1 & 0 & 0 & 0 & 0 & 0 \\
0 & 1 & -\frac{1}{2} & 0 & 0 & 0 \\
0 & -\frac{1}{2} & 1 & -\frac{1}{2} & 0 & 0 \\
0 & 0 & -\frac{1}{2} & 1 & 0 & 0 \\
0 & 0 & 0 & 0 & 1 & 0 \\
0 & 0 & 0 & 0 & 0 & 1 \end{array} \right)
\ee
\be
(\det{g})^{1\over2} = \frac{1}{\sqrt{2}}.
\ee
\item
$\underline{\bf \mbb{Z}_4, A_3\times A_3}$

Discussed in section \ref{a3a3section}.
\item
$\underline{\bf \mbb{Z}_6, A_2\times G_2\times G_2}$

Discussed in \cite{bgk99a}.
\item
$\underline{{\bf \mbb{Z}_6, G_2\times A_2\times A_2}, \Theta = \Gamma_1
  \Gamma_2 \Gamma_3 \Gamma_4 P_{36} P_{45}}$
\begin{figure}[ht]
\begin{center}
\makebox[10cm]{
\epsfxsize=15cm
\epsfysize=5cm
\epsfbox{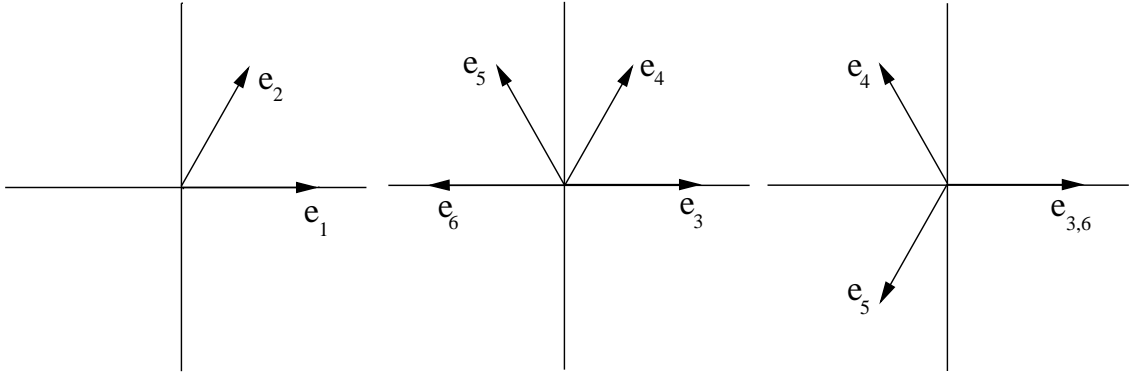}}
\end{center}
\caption{The $\mbb{Z}_6, G_2\times A_2\times A_2$ lattice}
\label{g2a2a2pic}
\end{figure}
%
\begin{eqnarray*}
\omega {\bf e}_1 = {\bf e}_2\qquad
\omega {\bf e}_2 &=& -{\bf e}_1 + {\bf e}_2\qquad
\omega {\bf e}_3 = {\bf e}_4\\
\omega {\bf e}_4 = {\bf e}_5\qquad
\omega {\bf e}_5 &=& {\bf e}_6\qquad
\omega {\bf e}_6 = -{\bf e}_3 - {\bf e}_5.
\end{eqnarray*}
\be g_{ij} = {\bf e}_i \cdot {\bf e}_j = \left( \begin{array}{cccccc}
1 & \frac{1}{2} & 0 & 0 & 0 & 0 \\
\frac{1}{2} & 1 & 0 & 0 & 0 & 0 \\
0 & 0 & 1 & 0 & -\frac{1}{2} & 0 \\
0 & 0 & 0 & 1 & 0 & -\frac{1}{2} \\
0 & 0 & -\frac{1}{2} & 0 & 1 & 0 \\
0 & 0 & 0 & -\frac{1}{2} & 0 & 1 \end{array} \right)
\ee
\be
(\det{g})^{1\over2} = \frac{3\sqrt{3}}{8}.
\ee
\item
$\underline{\bf \mbb{Z}'_6, A_1\times A_1\times A_2\times G_2}$

Discussed in \cite{bgk99a}.
\item
$\underline{\bf \mbb{Z}'_6, A_2\times D_4}$

For this orbifold there did not appear to be a natural description in
terms of the formalism outlined in section \ref{Amplitudes}.
\item
$\underline{{\bf \mbb{Z}'_6, A_1\times A_1\times A_2\times A_2}, \Theta
  = \Gamma_1 \Gamma_2 \Gamma_3 \Gamma_4 P_{36} P_{45}}$
\begin{figure}[ht]
\begin{center}
\makebox[10cm]{
\epsfxsize=15cm
\epsfysize=5cm
\epsfbox{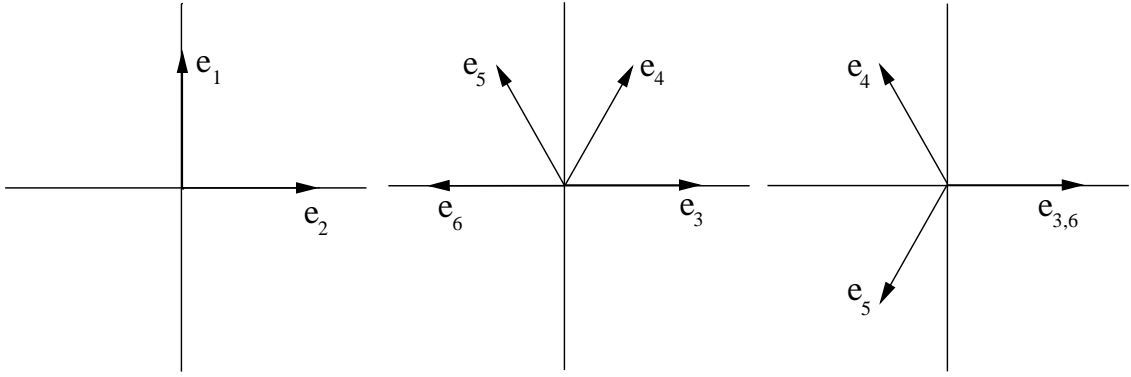}}
\end{center}
\caption{The $\mathbb{Z}'_6, A_1\times A_1\times A_2\times A_2$ lattice}
\label{a1a1a2a2}
\end{figure}
%
\begin{eqnarray*}
\omega {\bf e}_1 = -{\bf e}_1\qquad
\omega {\bf e}_2 &=& -{\bf e}_2\qquad
\omega {\bf e}_3 = {\bf e}_4\\
\omega {\bf e}_4 = {\bf e}_5\qquad
\omega {\bf e}_5 &=& {\bf e}_6\qquad
\omega {\bf e}_6 = -{\bf e}_3 - {\bf e}_5.
\end{eqnarray*}
\be g_{ij} = {\bf e}_i \cdot {\bf e}_j = \left( \begin{array}{cccccc}
1 & 0 & 0 & 0 & 0 & 0 \\
0 & 1 & 0 & 0 & 0 & 0 \\
0 & 0 & 1 & 0 & -\frac{1}{2} & 0 \\
0 & 0 & 0 & 1 & 0 & -\frac{1}{2} \\
0 & 0 & -\frac{1}{2} & 0 & 1 & 0 \\
0 & 0 & 0 & -\frac{1}{2} & 0 & 1 \end{array} \right)
\ee
\be
(\det{g})^{1\over2} = \frac{3}{4}.
\ee
\item
$\underline{\bf \mbb{Z}'_6, A_1\times A_5}$
\begin{figure}[ht]
\begin{center}
\makebox[10cm]{
\epsfxsize=15cm
\epsfysize=5cm
\epsfbox{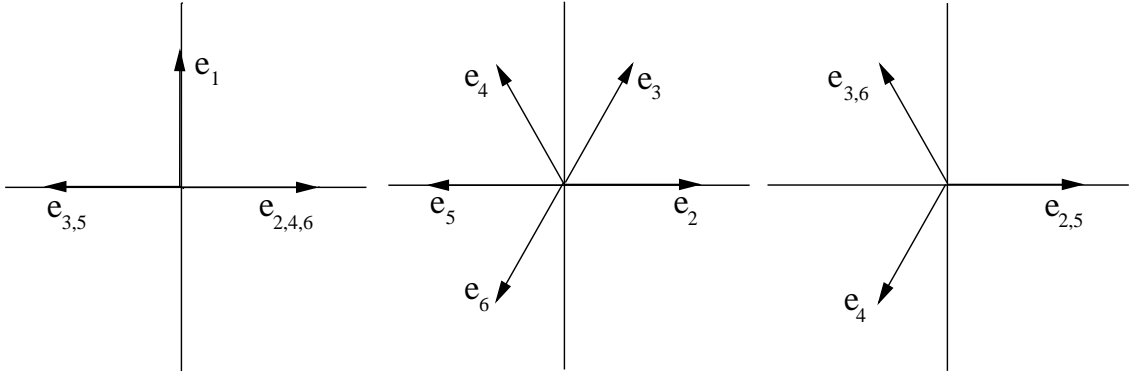}}
\end{center}
\caption{The $\mbb{Z}'_6, A_1\times A_5$ lattice}
\label{a1a5pic}
\end{figure}
%
\begin{eqnarray*}
\omega {\bf e}_1 = -{\bf e}_1\qquad
\omega {\bf e}_2 &=& {\bf e}_3\qquad
\omega {\bf e}_3 = {\bf e}_4\\
\omega {\bf e}_4 = {\bf e}_5\qquad
\omega {\bf e}_5 &=& {\bf e}_6\qquad
\omega {\bf e}_6 = -{\bf e}_2 - {\bf e}_3 - {\bf e}_4 - {\bf e}_5 - {\bf e}_6.
\end{eqnarray*}
\be g_{ij} = {\bf e}_i \cdot {\bf e}_j = \left( \begin{array}{cccccc}
1 & 0 & 0 & 0 & 0 & 0 \\
0 & 1 & -\frac{1}{2} & 0 & 0 & 0 \\
0 & -\frac{1}{2} & 1 & -\frac{1}{2} & 0 & 0 \\
0 & 0 & -\frac{1}{2} & 1 & -\frac{1}{2} & 0 \\
0 & 0 & 0 & -\frac{1}{2} & 1 & -\frac{1}{2} \\
0 & 0 & 0 & 0 & -\frac{1}{2} & 1 \end{array} \right)
\ee
\be
(\det{g})^{1\over2} = \frac{3}{4}.
\ee
\item
$\underline{\bf\mbb{Z}_7, A_6}$

See section \ref{z7section}.
\item
$\underline{\bf \mbb{Z}_8, B_4\times D_2}$

See section \ref{z8section}.
\item
$\underline{\bf \mbb{Z}_8, A_1\times D_5}$
\begin{figure}[ht]
\begin{center}
\makebox[10cm]{
\epsfxsize=15cm
\epsfysize=5cm
\epsfbox{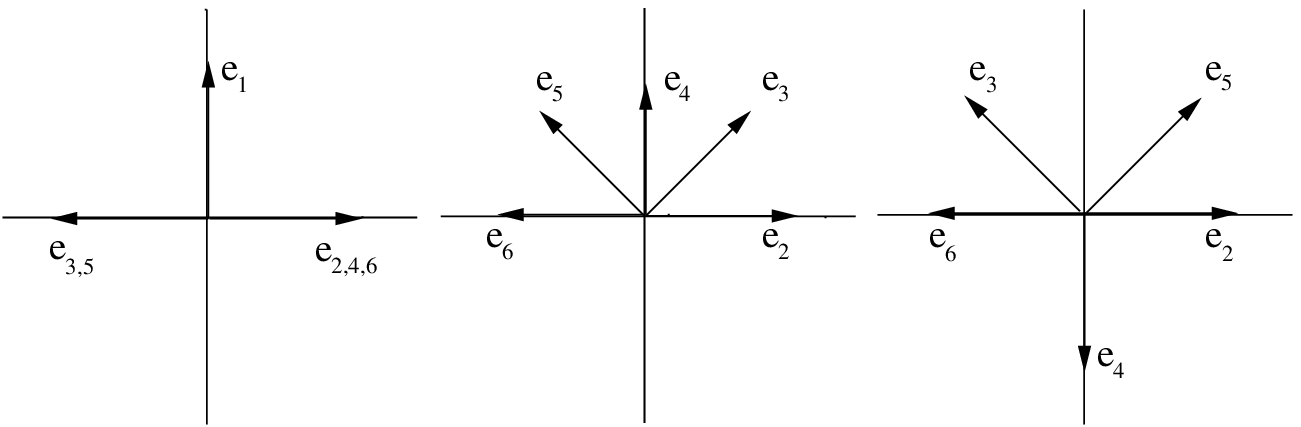}}
\end{center}
\caption{The $\mathbb{Z}_{8}, A_1\times D_5$ lattice}
\label{a1d5picture}
\end{figure}
%
\begin{eqnarray*}
\omega {\bf e}_1 = -{\bf e}_1\qquad
\omega {\bf e}_2 &=& {\bf e}_3\qquad
\omega {\bf e}_3 = {\bf e}_4\\
\omega {\bf e}_4 = {\bf e}_5\qquad
\omega {\bf e}_5 &=& {\bf e}_6\qquad
\omega {\bf e}_6 = -{\bf e}_2 - {\bf e}_3 - {\bf e}_6.
\end{eqnarray*}
\be g_{ij} = {\bf e}_i \cdot {\bf e}_j = \left( \begin{array}{cccccc}
1 & 0 & 0 & 0 & 0 & 0 \\
0 & 1 & -\frac{1}{2} & \frac{1}{2} & -\frac{1}{2} & 0 \\
0 & -\frac{1}{2} & 1 & -\frac{1}{2} & \frac{1}{2} & -\frac{1}{2} \\
0 & \frac{1}{2} & -\frac{1}{2} & 1 & -\frac{1}{2} & \frac{1}{2} \\
0 & -\frac{1}{2} & \frac{1}{2} & -\frac{1}{2} & 1 & -\frac{1}{2} \\
0 & 0 & -\frac{1}{2} & \frac{1}{2} & -\frac{1}{2} & 1 \end{array} \right)
\ee
\be
(\det{g})^{1\over2} = \frac{1}{2\sqrt{2}}.
\ee
\item
$\underline{\bf \mbb{Z}'_8, B_2\times B_4}$

See section \ref{z8psection}.
\item
$\underline{{\bf \mbb{Z}'_8, A_3\times A_3}, \Theta = \Gamma_1 \Gamma_2
  \Gamma_3  P_{16} P_{25} P_{34}}$
\begin{figure}[ht]
\begin{center}
\makebox[10cm]{
\epsfxsize=15cm
\epsfysize=5cm
\epsfbox{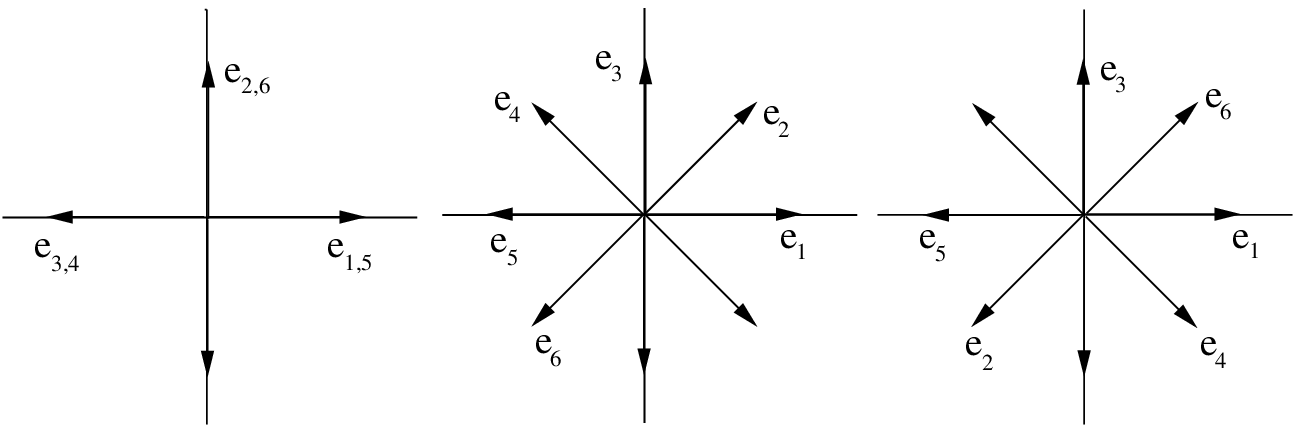}}
\end{center}
\caption{The $\mathbb{Z}'_8, A_3\times A_3$ lattice}
\label{a3a3picture}
\end{figure}
%
\begin{eqnarray*}
\omega {\bf e}_1 = {\bf e}_2\qquad
\omega {\bf e}_2 &=& {\bf e}_3\qquad
\omega {\bf e}_3 = {\bf e}_4\\
\omega {\bf e}_4 = {\bf e}_5\qquad
\omega {\bf e}_5 &=& {\bf e}_6\qquad
\omega {\bf e}_6 = -{\bf e}_1 - {\bf e}_3 - {\bf e}_5.
\end{eqnarray*}

\be g_{ij} = {\bf e}_i \cdot {\bf e}_j = \left( \begin{array}{cccccc}
1 & 0 & -\frac{1}{2} & 0 & 0 & 0 \\
0 & 1 & 0 & -\frac{1}{2} & 0 & 0 \\
-\frac{1}{2} & 0 & 1 & 0 & -\frac{1}{2} & 0 \\
0 & -\frac{1}{2} & 0 & 1 & 0 & -\frac{1}{2} \\
0 & 0 & -\frac{1}{2} & 0 & 1 & 0 \\
0 & 0 & 0 & -\frac{1}{2} & 0 & 1 \end{array} \right)
\ee
\be
(\det{g})^{1\over2} = {1\over2}.
\ee
\item
$\underline{\bf \mbb{Z}_{12}, A_2\times F_4}$

See section \ref{z12section}.
\item
$\underline{\bf \mbb{Z}_{12}, E_6}$
\begin{figure}[ht]
\begin{center}
\makebox[10cm]{
\epsfxsize=15cm
\epsfysize=5cm
\epsfbox{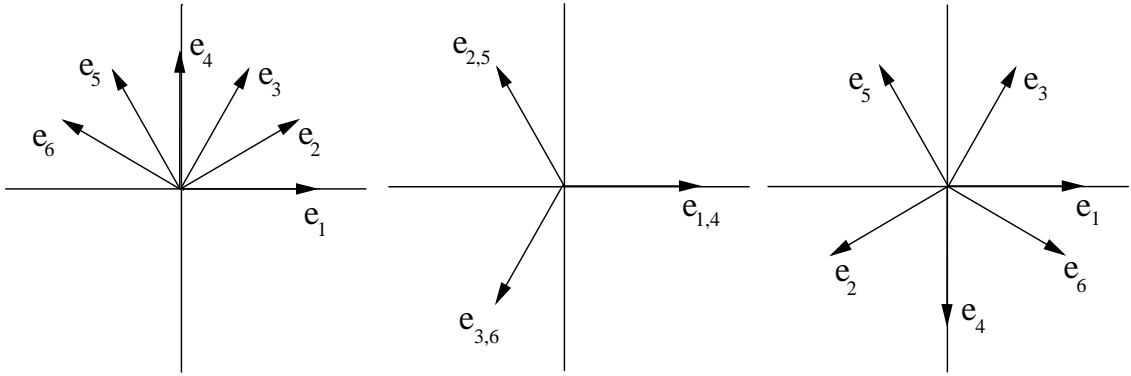}}
\end{center}
\caption{The $\mathbb{Z}_{12}, E_6$ lattice}
\label{e6picture}
\end{figure}
%
\begin{eqnarray*}
\omega {\bf e}_1 = {\bf e}_2,\qquad
\omega {\bf e}_2 &=& {\bf e}_3,\qquad
\omega {\bf e}_3 = {\bf e}_4\\
\omega {\bf e}_4 = {\bf e}_5,\qquad 
\omega {\bf e}_5 &=& {\bf e}_6,\qquad
\omega {\bf e}_6 = {\bf e}_4 - {\bf e}_1 - {\bf e}_2-{\bf e}_6.
\end{eqnarray*}
\be g_{ij} = {\bf e}_i \cdot {\bf e}_j = \left( \begin{array}{cccccc}
1 & -\frac{1}{2} & 0 & \frac{1}{2} & -\frac{1}{2} & 0 \\
-\frac{1}{2} & 1 & -\frac{1}{2} & 0 & \frac{1}{2} & -\frac{1}{2} \\
0 & -\frac{1}{2} & 1 & -\frac{1}{2} & 0 & \frac{1}{2} \\
\frac{1}{2} & 0 & -\frac{1}{2} & 1 & -\frac{1}{2} & 0 \\
-\frac{1}{2} & \frac{1}{2} & 0 & -\frac{1}{2} & 1 & -\frac{1}{2} \\
0 & -\frac{1}{2} & \frac{1}{2} & 0 & -\frac{1}{2} & 1 \end{array} \right)
\ee
\be
(\det{g})^{1\over2} = {\sqrt{3}\over8}.
\ee
\item
$\underline{\bf \mbb{Z}'_{12}, D_2\times F_4}$

See section \ref{z12psection}.
\end{enumerate}

\end{appendix}

\clearpage
\bibliography{rev}
\bibliographystyle{utphys}

\end{document}